\documentclass[12pt]{article}
\usepackage[T1]{fontenc}
\usepackage{lmodern}
\usepackage[margin=1.25in]{geometry}
\usepackage{amsmath,amssymb,amsthm}
\usepackage{booktabs}
\usepackage{graphicx}
\usepackage{enumitem}
\usepackage{setspace}
\usepackage[authoryear,round]{natbib}
\usepackage[colorlinks=true,linkcolor=black,citecolor=black,urlcolor=black]{hyperref}
\newtheorem{theorem}{Theorem}
\newtheorem{proposition}{Proposition}

\newtheorem{remark}{Remark}
\newtheorem{corollary}{Corollary}
\newcommand{\R}{\mathbb{R}}
\newcommand{\F}{\mathbb{F}}
\newcommand{\Z}{\mathbb{Z}}
\providecommand{\C}{}\renewcommand{\C}{\mathbb{C}}
\renewcommand{\H}{\mathrm{H}}
\newcommand{\econ}{\textbf{Economic statement.}\ }
\newcommand{\formal}{\par\smallskip\textbf{Formal statement.}\ }

\title{\textbf{Links Are Not Enough: Higher-Order Structure and Bargaining over Coherent Agreements}}
\author{Tomohiro Uchiyama\thanks{Faculty of International Liberal Arts, Soka University, Tokyo, Japan. Email: t.uchiyama2170@gmail.com}}
\date{}

\begin{document}
\maketitle

\begin{abstract}
\noindent Pairwise agreements, from tax treaties to correspondent banking, can contradict one another: routing through a third party can beat dealing directly. Coherence depends on triples, not links, so graph models miss it. We introduce the \emph{coherence gap}, a higher-order spectral gap of three-party checks, and show that decentralized agreement-making fails to produce it: members of checks repairing global inconsistencies capture about $3/n$ of their value among $n$ symmetric countries. Our solution is a package of these checks that enters into force only if every participant ratifies. Higher-order structure decides who must join: a coalition can repair only inconsistencies running through its own links, and a country whose agreements leave its partners disconnected is indispensable; in equilibrium bargaining, such countries split the whole surplus. Using trade-agreement data on 80 economies, we show that the EU cannot repair the network alone, while Ukraine and Costa Rica can block its repair.

\medskip
\noindent\textbf{Keywords:} international integration; multilateral agreements; network formation; higher-order networks; public goods; spectral gap.\\
\textbf{JEL codes:} D85, C71, C78, F15, F53.
\end{abstract}

\section{Introduction}

Suppose three countries accept one another's product certificates. Every pair has an agreement, yet a firm may still need a different test at each border, because the three bilateral promises were never reconciled. The same problem arises among banks with different payment rules, firms with overlapping technical standards and countries with bilateral tax treaties, where routing a payment through a third country can be cheaper than paying directly.

This paper shows that whether bilateral agreements fit together is a higher-order property, and that this higher-order structure shapes classical economic questions about them: who builds the agreements that make a system coherent, who must take part, and who can hold them up. A network of bilateral links records who is connected to whom; whether the agreements are consistent depends on three-party checks that verify that a triple of arrangements fits together. We first develop the framework and show what coherence requires. We then show that decentralized agreement-making fails to produce it, and that the economics of producing it instead, through packages, coalitions and bargaining, is governed by the topology of the checks.

\textbf{The framework.} We define the \emph{coherence gap} $\Gamma$, a higher-order spectral gap of the checks, and show that the worst-case persistence of disturbances is $1/(\kappa\Gamma)$, where $\kappa$ is the capacity of those who correct them (Theorem~\ref{thm:1}). The correction rule is what competitive arbitrage produces when a check makes a three-country round trip tradeable, and each country's welfare loss is its \emph{exposure}, the variance of inconsistency on its own links (Proposition~\ref{prop:micro}). A growing mesh of local checks loses the gap (Theorem~\ref{thm:2}), while sparse architectures built from high-dimensional expanders keep it at linear cost (Theorem~\ref{thm:3} and Proposition~\ref{prop:local}).

\textbf{The formation wedge.} A new check lowers the exposure of every country whose links it disciplines, and its members capture only the part that falls on their own links (Proposition~\ref{prop:wedge}). This share is small exactly when the network is weak and its weak directions are spread over many countries, about $3/n$ among $n$ symmetric countries (Theorem~\ref{thm:general}). Once a coherent architecture exists, members capture a uniformly positive share of further checks (Corollary~\ref{cor:expanders}). Decentralization can maintain coherence but need not reach it.

\textbf{Bargaining over coherence.} The main results concern how coherence can be reached instead, and each turns on higher-order structure. First, a package of checks negotiated together can enlist the countries on whose links the value falls, but only if it enters into force all-or-nothing: if it applies among those who ratify, a participant can refuse and keep the benefit of the others' checks, and a package of checks that substitute for one another is no easier to finance than its checks one by one (Theorem~\ref{thm:package}). Second, a coalition can strengthen only the directions of inconsistency that run through links among its own members, so the identity of the participants matters more than their number (Proposition~\ref{prop:table}). Third, some countries are indispensable. If a country's partners fall into groups that no common agreement with it joins, no coalition without it can make the economy coherent; more generally, the gap reachable without a country is bounded above by the connectivity of its partner graph, and, when the other countries are coherent among themselves, bounded below by a fixed fraction of it (Theorem~\ref{thm:veto}). This is the converse of the local-to-global principle behind expanders: well-connected partner graphs everywhere guarantee coherence, and one poorly connected partner graph caps it. Finally, in the cooperative game of reaching a coherence target, the core is empty for modest targets and, for ambitious ones, gives the whole surplus to the indispensable countries (Corollary~\ref{thm:core}). In the bargaining protocol of \citet{baronferejohn1989}, patient countries reach exactly this outcome: the indispensable countries split the surplus equally and everyone else receives nothing (Theorem~\ref{thm:bargain}). Because the gap is set by the weakest direction, coherence is highly complementary: when the weak directions run through many countries' links, their marginal contributions add up to many times the total gain, so if each can claim even a small share of what it adds, no division satisfies them all (Proposition~\ref{prop:holdup}). Hold-up power over coherence comes from the local shape of a country's agreements, not from its size.

\textbf{Evidence.} In the 80-economy trade-agreement network in DESTA \citep{dur2014,desta2025}, the bilateral graph is well connected and highly clustered, yet 76 independent directions of inconsistency are unchecked, and the members of the checks that would repair them would capture only about 15 percent of their value. Accessions since 1980 follow the value members keep, not the value they create, a pattern carried by which triples are checked rather than by who is linked. Built as one all-or-nothing package, the targeted checks' 47 participants would capture 92 percent of its value; under partial entry, they would pay for at most 14 percent. The 28 economies of the EU-28, adding every check among themselves, would not raise the gap at all, while a coalition of 27 mostly non-EU economies raises it almost a hundredfold. Without Ukraine or Costa Rica, no set of checks can raise the gap above 0.08, against 0.97 with all economies,\footnote{Without Germany or France, it still reaches 0.97. The ten most indispensable economies receive half of the Shapley value of the gains from coherence (Appendix~\ref{S-app:shapley}).} and the marginal contributions of the 80 economies sum to 17 times the total gain.

\textbf{Why existing tools are not enough.} Network economics represents interactions as a graph of bilateral links and studies how it shapes incentives, public goods and the propagation of shocks \citep{jackson2008,ballester2006,bramoulle2007,bramoulle2014,elliott2014,acemoglu2015}. Coherence is not a property of the graph: two economies can have the same bilateral graph and different checks, and no statistic of the graph separates them. Strategic network formation \citep{jacksonwolinsky1996,blochjackson2007} and the literature on networks of trade agreements \citep{goyaljoshi2006,furusawakonishi2007} study links whose value to outsiders comes from the paths they create; a check creates no path, and existing results do not say how its non-local value is split between members and outsiders. Cooperative games on networks restrict which coalitions can form to those connected in a graph \citep{myerson1977}; here what a coalition can do is restricted by the higher-order structure, through the triangles and cycles its members' links span. Mathematics and computer science supply architectures with a large higher-order gap \citep{lubotzky2018,dinur2017,panteleev2022} but say nothing about who would build them or who could block them.\footnote{\citet{elliottgolubleduc2022} show that individually rational choices can produce a fragile network; we identify a different reason, the spectral spread of a check's benefits. The treaty literature measures the content and depth of agreements \citep{dur2014,baccini2015,mattoo2020}, not whether they are consistent with one another.}

\textbf{What is new.} The paper brings into economics the cohomology of simplicial complexes, Garland's local-to-global method and the Ramanujan complexes of \citet{lsv2005a,lsv2005b} \citep{lubotzky2018}, and uses them to identify the coherence gap as the object that governs how long inconsistencies persist. The formation wedge, the results on package agreements, coalitions and indispensable countries, the equilibrium bargaining and hold-up results, and the identification of who holds hold-up power over coherence are new, as is the empirical work: a measurement of the higher-order coherence of a real network of agreements, a test of the formation wedge on observed decisions, and the identification of the countries that hold up its repair.

\section{Framework: bilateral adjustments and multilateral checks}\label{sec:framework}

There is a set $V$ of $n$ participants (countries, banks, firms) and a connected set $E$ of $m$ bilateral relationships, or \emph{links}. Each relationship $ij$ carries an \emph{adjustment} $x_{ij}=-x_{ji}$, the amount needed to translate between $i$'s and $j$'s conventions: an exchange rate between two benchmarks, a conversion factor between two standards, or the premium of one jurisdiction's certification over another's. The system is \textbf{globally consistent} if each participant has a benchmark value $p_i$ with
\begin{equation}\label{eq:1}
x_{ij}=p_j-p_i \quad\text{for every relationship } ij.
\end{equation}
This is the no-arbitrage condition: translating from $i$ to $j$ to $k$ and back returns you to the start.

A \textbf{check} is a three-party procedure for a triple $ijk$ whose three links exist. It verifies that the adjustments around the triangle cancel,
\begin{equation}\label{eq:2}
x_{ij}+x_{jk}+x_{ki}=0.
\end{equation}
Let $T$ be the set of checks, and call $X=(V,E,T)$ the \emph{network}. Global consistency implies that every check passes; the design question is the converse. A choice of benchmarks is not an inconsistency, so we study the \emph{inconsistent part} of the adjustments, the part that no benchmark assignment explains. These parts form a space $C$ of dimension $r=m-n+1$, the number of independent cycles of the bilateral network.\footnote{Formally, adjustments are vectors in $\R^E$, $d_0:\R^V\to\R^E$ maps benchmarks to adjustments, and $C=(\operatorname{im}d_0)^\perp$; $D$ is the map $d_1:\R^E\to\R^T$ from adjustments to check residuals, restricted to $C$ (Appendix~\ref{app:proofs}).}

\textbf{The coherence gap.} Each inconsistency $z\in C$ produces a vector of check residuals $Dz$. The \textbf{coherence gap} is the smallest total squared signal that any unit-size inconsistency produces:
\begin{equation}\label{eq:3}
\Gamma(X)=\min_{z\in C,\ \|z\|=1}\|Dz\|^2=\lambda_{\min}(H),\qquad H=D^*D,
\end{equation}
If $\Gamma=0$, some inconsistency is invisible to every check. If $\Gamma$ is positive but small, every inconsistency is detectable but some produce only weak signals. Here $D^*$ is the adjoint of $D$, so $H$ is a symmetric nonnegative operator on $C$. The distinction between the last two cases matters as the economy grows: for a family of economies indexed by size $n$, $\Gamma_n$ can be positive at every size and still shrink to zero, and we say that a family of economies is \emph{coherent} if $\Gamma_n$ is bounded away from zero.

\textbf{Invisible inconsistencies.} Call adjustments \emph{locally consistent} if every check passes. The first cohomology group $\H^1(X;\R)$, locally consistent adjustments modulo globally consistent ones, counts the independent ways the agreements can contradict one another while every check passes. We call these \emph{uncovered cycles}; $\Gamma=0$ exactly when $\dim\H^1>0$.\footnote{The theory uses real coefficients. In the data we compute $\dim\H^1$ over the two-element field $\F_2$, where elimination is exact and fast; in general the two dimensions can differ, but in our data they coincide in every year from 1990 to 2022 in which we compute both.}

In a connected network of $m$ bilateral agreements among $n$ participants, $\dim\H^1\ge m-n+1-|T|$, so at least $m-n+1$ independent checks are needed before every inconsistency can be detected, and each new link that no check covers adds one uncovered cycle (Proposition~\ref{S-prop:1}).

Bilateral integration therefore raises the burden on multilateral institutions. Adding a link can even improve the connectivity of the bilateral graph while breaking the certification of global consistency.\footnote{With four participants, links $12,13,14,23,34$ and checks on $123$ and $134$, every cycle is checked and the graph Laplacian has eigenvalues $(0,2,4,4)$. Adding link $24$ without a check raises them to $(0,4,4,4)$ but creates a cycle that no check detects.}

\section{Coherence and the persistence of disturbances}\label{sec:correction}

\subsection{Correction}

Detecting a contradiction is not the same as correcting it. Each check $\tau$ is run by an intermediary (a regulator, certifier or correspondent bank) that observes the residual $(Dz)_\tau$, is rewarded in proportion to the discrepancy it resolves, and chooses effort $y_\tau$ at cost $y_\tau^2/(2\kappa)$, where the \emph{response capacity} $\kappa>0$ measures staff or budget. This gives $y=\kappa Dz$. Correction moves the inconsistency by $-D^*y$, and shocks $\xi(t)\in C$ keep creating new mismatches, so
\begin{equation}\label{eq:5}
\dot z(t)=-\kappa H z(t)+\xi(t),\qquad z(0)=0.
\end{equation}
Without shocks, the weakest-signalled direction decays at rate $\kappa\Gamma$, so capacity and architecture are perfect substitutes. We measure performance by the worst-case long-run amplification of unit-bounded disturbances,
\begin{equation}\label{eq:6}
A(X)=\sup_{\|\xi\|_{L^\infty}\le 1}\ \limsup_{t\to\infty}\|z(t)\|.
\end{equation}

\begin{theorem}[Persistence of disturbances]\label{thm:1}
\econ The worst-case long-run size of a persistent disturbance is one divided by the product of response capacity and the coherence gap. If some inconsistency is invisible to every check, disturbances grow without bound. A growing economy keeps disturbances bounded if and only if it is coherent.
\formal Under \eqref{eq:5}, $A(X)=1/(\kappa\Gamma(X))$ if $\Gamma(X)>0$ and $A(X)=+\infty$ if $\Gamma(X)=0$.
\end{theorem}

The theorem separates a \emph{logical failure}, $\Gamma=0$, from a \emph{quantitative failure}, a small positive $\Gamma$, which is the main concern in large systems.\footnote{The constant $1/(\kappa\Gamma)$ relies on linear, independent intermediaries, but the scaling conclusions do not. With nonlinear responses or coordinated intermediaries, or with heterogeneous capacities, amplification is finite exactly when $\Gamma>0$ and is bounded above and below by multiples of $1/\Gamma$ (Proposition~\ref{S-prop:robust} in the Supplemental Appendix).}

\subsection{An arbitrage foundation}\label{sec:micro}

The reward for resolving discrepancies, the impact of effort and the welfare measure of Section~\ref{sec:formation} all follow from a market model in which a check is a route that can be traded. Each link $e=ij$ hosts a market for cross-border transactions in which $x_e$ is a relative price. Demand and supply are linear, so a deviation $z_e$ from a consistent benchmark causes a deadweight loss $\frac\eta2z_e^2$ per unit of time, borne equally by the two endpoints; $\eta>0$ is the depth of the market. For each checked triple $\tau$, competitive arbitrageurs can execute round trips $i\to j\to k\to i$; the check provides the common recognition or clearing without which the three legs cannot be combined, and round trips through unchecked triples are not available. A round trip of size $y_\tau$ earns $y_\tau(Dz)_\tau$, costs $y_\tau^2/(2\kappa)$ in capital charges, and moves the price of each of its links by one unit per unit traded \citep[linear price impact, as in][]{kyle1985}. Shocks arrive as white noise $\sigma\,dW$, with $W$ a standard Brownian motion on $C$ and $\sigma>0$.

\begin{proposition}[Arbitrage foundation]\label{prop:micro}
\econ Competitive arbitrage around checked triples produces exactly the correction rule \eqref{eq:5}. The deadweight loss each country bears is proportional to its exposure, the variance of inconsistency on its links, while total arbitrage rents do not depend on the architecture. Exposure is therefore the right welfare measure for comparing architectures.
\formal (a) Arbitrage flows are $y=\kappa Dz$, and $dz=-\kappa Hz\,dt+\sigma\,dW$. (b) If $\Gamma>0$, the stationary covariance of $z$ is $\Sigma=\frac{\sigma^2}{2\kappa}H^{-1}$, and country $i$'s expected deadweight loss per unit of time is $\frac\eta4L_i$, where
\begin{equation}\label{eq:exposure}
L_i=\sum_{e\ni i}\Sigma_{ee}
\end{equation}
is its \emph{exposure}; we write $L_i(T)$ when the dependence on the checks $T$ matters. (c) Expected arbitrage rents equal $\frac{\sigma^2}{4}r$ for every architecture with $\Gamma>0$. (d) Adding a check $\tau$ raises the rents earned on $\tau$ from zero to $\frac{\sigma^2}{4}\frac{q}{1+q}<\frac{\sigma^2}{4}$, with $q$ defined in Section~\ref{sec:formation}, and lowers the rents on other checks by the same amount.
\end{proposition}

Here $\kappa$ is the arbitrage sector's risk-bearing capacity, the object that limits arbitrage in \citet{grombvayanos2002}. Since rents do not depend on the architecture, architectures differ in welfare only through total exposure; by part (d), members who own a check's arbitrage business gain a business-stealing rent of at most $\sigma^2/4$.\footnote{When adjustments are set by statute rather than by trading, a round trip has no price impact, and arbitrage earns rents without removing the discrepancy. Treaty shopping through bilateral tax treaties is an example (Appendix~\ref{S-app:tax}): routing gains persist because withholding rates move only when treaties are renegotiated. Correction then has to come from the intermediaries above, and the gap governs their task in the same way.}

\section{Local meshes and expanders}\label{sec:arch}

\subsection{Local checks deteriorate with scale}

Can a modest number of local checks keep an economy coherent as it grows? Take a triangular mesh, in which neighbouring participants are linked and each local triple has a check. Every inconsistency is detected, yet the system is weak against disturbances spread across many cells.

\begin{theorem}[Local agreements do not scale]\label{thm:2}
\econ If agreements and checks are arranged purely locally, the coherence gap shrinks as the economy grows, and persistent disturbances are amplified without bound, even though every inconsistency is detectable and the number of agreements grows only in proportion to the economy.
\formal If $X$ triangulates a topological disk, with $t$ triangular checks and $b_\partial$ links on its boundary, then $0<\Gamma(X)\le b_\partial/t$. For a triangular patch of side $L$, with $n=(L+1)(L+2)/2$ participants, $\Gamma\le3/L$ and $A\ge L/(3\kappa)$.
\end{theorem}

A loop of small adjustments circulating around a large region is almost invisible: adjacent contributions offset one another on shared links, and only the boundary registers the imbalance.

\subsection{Sparse architectures that scale}

Call a family of architectures a \emph{high-dimensional expander} if every link lies in a bounded number of checks and $\Gamma$ stays bounded away from zero. By Theorem~\ref{thm:1}, at bounded local complexity a family is coherent exactly when it has this property, so the requirement is a definition rather than a further theorem; dense architectures, in which links lie in unboundedly many checks, can be coherent without being expanders in this sense. Theorem~\ref{thm:2} shows that one natural family, local meshes, fails the requirement; it does not show that every alternative arrangement does. Explicit examples come from Ramanujan complexes \citep{lsv2005a,lsv2005b,lubotzky2018}.

\begin{theorem}[Sparse architectures that scale]\label{thm:3}
\econ There are architectures in which each participant has a bounded number of agreements, the numbers of agreements and checks are proportional to the number of participants, and the coherence gap stays above a fixed constant. Amplification then stays bounded at total cost proportional to the size of the economy.
\formal There exist connected two-dimensional networks $Y_n$ with $|V(Y_n)|\to\infty$, bounded degrees and $\Gamma(Y_n)\ge9-4\sqrt2$, namely suitable finite quotients of the Bruhat--Tits building of $\mathrm{PGL}_3$ over the field $\F_8((t))$ of formal Laurent series over the eight-element field $\F_8$. They have $|E(Y_n)|=73|V(Y_n)|$ links and $|T(Y_n)|=219|V(Y_n)|$ checks, and $\sup_nA(Y_n)\le1/(\kappa(9-4\sqrt2))$.
\end{theorem}

Both families are sparse, but only one keeps its gap. The proof is local to global: Garland's spectral inequality, applied to each participant's network of partners, delivers a gap for the whole system, and each participant can run the same test on its own neighbourhood. For a participant $v$, let its \emph{partner graph} $\mathrm{Lk}(v)$ join two of its partners when $v$ and both of them share a check, and call $v$ \emph{$\lambda$-sound} if $\mathrm{Lk}(v)$ is connected and the second-smallest eigenvalue of its normalized Laplacian is at least $\lambda$.

\begin{proposition}[Local soundness]\label{prop:local}
\econ If every participant is $\lambda$-sound for some $\lambda>1/2$, the system is coherent whatever its size: every inconsistency is exposed with strength at least $2\lambda-1$.
\formal Let $X$ be connected with every link in at least $t_{\min}\ge1$ checks, and every participant $\lambda$-sound for some $\lambda>1/2$. Then $\dim\H^1(X;\R)=0$ and $\Gamma(X)\ge t_{\min}(2\lambda-1)$. If every participant has at most $\bar d$ partners, $|E|\le\bar dn/2$ and $|T|\le\bar d^2n/6$.
\end{proposition}

A triangular mesh sits exactly on the boundary: an interior node's partners form a six-cycle, whose second eigenvalue is exactly $1/2$.

\subsection{The cost of a target}

Architecture also substitutes for response capacity. To keep the worst-case persistence below a target, a local mesh needs capacity per check that grows with the square root of the economy's size, so its total cost grows like $n^{3/2}$, while the sparse architecture needs constant capacity per check and a total cost proportional to the economy (Proposition~\ref{S-prop:cost}).

Architectural quality economizes on response capacity, and its value rises with the size of the system.\footnote{The Supplemental Appendix shows that regional blocs whose checks stop at their borders leave at least one unchecked direction per cross-bloc link, net of the number of blocs (Proposition~\ref{S-prop:3}).} Coherence therefore requires checks that tie distant parts of the network together. Such checks have diffuse benefits, which is why they are hard to build.

\section{Who builds the checks?}\label{sec:formation}

Coherence requires non-local checks. This section shows that the checks that matter most for the system are the ones their members value least.

\subsection{A formation game}

Fix the bilateral network and an initial architecture $T_0$ with $\Gamma(T_0)>0$. A \emph{candidate} is a triple $\tau$ whose three links exist and that is not yet checked. Creating the check costs $c>0$, which its three members can divide among themselves as they like, as in \citet{blochjackson2007}; outsiders cannot contribute. Each country's payoff is minus its exposure $L_i(T)$, defined in \eqref{eq:exposure} with the normalization $\sigma^2/(2\kappa)=1$; by Proposition~\ref{prop:micro} this is its deadweight loss up to scale. An architecture $T\supseteq T_0$ is \emph{stable} if no three countries would jointly gain from creating a check and no members of an added check would jointly gain from dropping it:
\begin{align*}
\textstyle\sum_{i\in\tau}\big[L_i(T)-L_i(T\cup\{\tau\})\big]&<c\qquad\text{for every candidate }\tau\notin T,\\
\textstyle\sum_{i\in\tau}\big[L_i(T\setminus\{\tau\})-L_i(T)\big]&\ge c\qquad\text{for every added check }\tau\in T\setminus T_0 .
\end{align*}
This is pairwise stability \citep{jacksonwolinsky1996} applied to three-party checks. The planner minimizes $\sum_iL_i(T)+c\,|T\setminus T_0|$. Our results concern the first condition: whether a check would be added.

For a candidate $\tau$ with boundary vector $b=b_\tau\in C$ (one entry $\pm1$ on each of its three links, with the orientation of \eqref{eq:2}), let $u=H^{-1}b$ and $q=q_\tau=b^\top u$. The vector $u$ is the check's \emph{response profile}: how strongly it moves the inconsistency on each link. Let $\nu_e(\tau)\in\{0,1,2\}$ be the number of members of $\tau$ that are endpoints of $e$.

\begin{proposition}[The formation wedge]\label{prop:wedge}
\econ A new check lowers every country's exposure, including the exposure of countries outside it. Its members capture the share of its total value that falls on their own links, and build it only if that share, times the value, covers the cost.
\formal Adding $\tau$ lowers country $i$'s exposure by $\frac1{1+q}\sum_{e\ni i}u_e^2\ge0$. The social value of $\tau$ is $S(\tau)=2\|u\|^2/(1+q)$, and the members' share is
\begin{equation}\label{eq:share}
s(\tau)=\frac{\sum_e\nu_e(\tau)\,u_e^2}{2\|u\|^2}\ \in[0,1].
\end{equation}
The members create $\tau$ if and only if $s(\tau)S(\tau)\ge c$; the planner adds it if and only if $S(\tau)\ge c$. Every candidate with $c\le S(\tau)<c/s(\tau)$ is worth adding but is not created.
\end{proposition}

The externality differs from the usual one in network formation, where a link benefits outsiders through the paths it creates. A check creates no path; it disciplines adjustments wherever its response profile reaches. Transfers confined to those directly involved cannot internalize it \citep{blochjackson2007}. A multilateral institution that can levy contributions from all countries can. Let $g_i(\tau)=L_i(T)-L_i(T\cup\{\tau\})$ be country $i$'s fall in exposure from $\tau$, so that $\sum_ig_i(\tau)=S(\tau)$. If each country pays the share $g_i(\tau)/S(\tau)$ of the cost, a Lindahl price for the check \citep{foley1970}, every country weakly gains exactly when $S(\tau)\ge c$, so the institution approves precisely the checks that raise total welfare; the contributions depend only on the public architecture, so no country gains by misreporting. This is the single-check case of Theorem~\ref{thm:package}. With voluntary contributions, outsiders would free-ride; Section~\ref{sec:package} shows how a package of checks can do without them.

\subsection{When is the wedge large?}

Two features of the network matter: how weak it is, and how its weak directions are spread across countries. Let $\varphi_1$ be a unit eigenvector of $H$ for $\Gamma$, and $\beta_\tau=|\varphi_1^\top b|$, the check's leverage on the weakest direction.

\begin{theorem}[The formation wedge in general networks]\label{thm:general}
\econ In a well-connected network, members capture a large share of any check they could build. The share can be small only if the network is weak and its weak directions are spread over many countries. When one weak direction dominates, the members of a check aimed at it capture roughly that direction's weight on their own links.
\formal For every architecture with $\Gamma>0$ and every candidate $\tau$:
\begin{enumerate}[nosep]
\item[(a)] $s(\tau)S(\tau)\le\frac{q}{1+q}\sum_{i\in\tau}L_i(T)$, $\ S(\tau)\ge\frac{2\beta_\tau^2}{\Gamma(\Gamma+3)}$, $\ s(\tau)S(\tau)\ge\frac{2q^2}{3(1+q)}$ and $s(\tau)\ge\frac{\Gamma q}3$.
\item[(b)] Suppose $\Gamma$ is a simple eigenvalue, let $\lambda_2$ be the next eigenvalue of $H$, and give each country the weight $\ell_i=\frac12\sum_{e\ni i}\varphi_1(e)^2$, so that $\sum_i\ell_i=1$. Let $s_1(\tau)=\sum_{i\in\tau}\ell_i$. If $\beta_\tau>0$, then
\[
\Big(\big(\sqrt{(1-\varepsilon^2)\,s_1(\tau)}-\varepsilon\big)^+\Big)^2\ \le\ s(\tau)\ \le\ \big(\sqrt{s_1(\tau)}+\varepsilon\big)^2,\qquad \varepsilon^2\le\frac{\Gamma}{\lambda_2}\Big(\frac{\Gamma q}{\beta_\tau^2}-1\Big),
\]
where $\varepsilon=\varepsilon(\tau)$ is the share of the norm of the response profile that lies outside the weakest direction.
\end{enumerate}
\end{theorem}

By (a), the share is at least $\Gamma q/3$, so the wedge requires a weak network, and a check acting on the weakest direction is worth more the smaller the gap. Part (b) shows that it also requires spread. The ratio $\Gamma q/\beta_\tau^2\ge1$ measures how well a check is aimed: it equals one when all of the check's leverage works through the weakest direction. For a well-aimed check, the members capture approximately $s_1(\tau)$, the combined weight of the weakest direction on their links. If the weights are spread over $n_{\mathrm{eff}}=1/\sum_i\ell_i^2$ countries, a typical triple holds about $3/n_{\mathrm{eff}}$ of them.

Theorem~\ref{thm:general} says when decentralized formation fails, and Section~\ref{sec:arch} says what success requires: at bounded local complexity, an economy stays coherent as it grows exactly when its checks form a high-dimensional expander in the sense of Section~\ref{sec:arch}. The economic question is therefore who builds coherent architectures.

\begin{corollary}[Incentives in coherent architectures]\label{cor:expanders}
\econ Once a coherent architecture exists, the members of any further check capture at least a fixed share of its value. Decentralized formation then builds every sufficiently valuable check. In the Ramanujan architectures of Theorem~\ref{thm:3}, members capture at least 12 percent of any check's value, whatever the size of the economy.
\formal Let every link of $T$ lie in at most $\bar t$ checks and let $\Gamma(T)\ge\gamma_0>0$. Then every candidate has $s(\tau)\ge\gamma_0/(3\bar t)$, and every candidate with $S(\tau)\ge3\bar t\,c/\gamma_0$ is created. For the architectures $Y_n$ of Theorem~\ref{thm:3}, $\bar t=9$ and $\gamma_0=9-4\sqrt2$, so $s(\tau)\ge0.12$.
\end{corollary}

The bound depends on the gap, not on the size of the economy. In the two benchmarks below, the gap vanishes as the economy grows and so does the members' share of the checks that matter. Once a coherent architecture exists, sufficiently valuable additions are privately sustainable, but decentralized formation need not reach coherence: the wedge is an obstacle to reaching coherence, not to maintaining it, and this is when coordination is needed. The corollary describes incentives once an expander exists; Section~\ref{sec:package} shows how the coordinated step that builds one can be taken.

\subsection{Two benchmarks}\label{sec:grid}

Two benchmark economies show that the wedge can grow without bound (Appendix~\ref{S-app:bench}). On a grid of local agreements, a missing check is worth about $(3/\pi)\log L$ in an $L\times L$ economy while its members' gain stays bounded, so for any cost above that bound the local architecture is stable but not efficient once the economy is large (Proposition~\ref{S-prop:grid}). When no country is special, as in a rotation-invariant economy on $\Z_n$, the weak directions are waves running through the whole economy, and an uncertainty principle shows that the members of a check that repairs a global inconsistency capture about $3/n$ of its value (Proposition~\ref{S-prop:symm}); this is the case $\ell_i=1/n$ of Theorem~\ref{thm:general}(b).

\begin{remark}[Invisible versus weak inconsistencies]\label{rem:invisible}
The wedge concerns small positive gaps. If $\Gamma=0$, white-noise shocks make the variance of $z$ grow linearly along $\ker H$. A check with a component in $\ker H$ removes a null direction that is nonzero on at least one of its own links, so the exposure of at least two members grows strictly more slowly, and over a long enough horizon their gain exceeds any cost. Countries therefore have a private incentive to make every inconsistency visible, but not to make it visible strongly.
\end{remark}

The 2022 treaty network is far from both benchmarks, but part (b) is informative there. Starting from the architecture of Section~\ref{sec:incentives}, the weakest direction carries 92 percent of the squared response of the median check in the top decile of Table~\ref{S-tab:shares}. Its weights are spread over $n_{\mathrm{eff}}=24.8$ countries, so $3/n_{\mathrm{eff}}=0.12$, and the median top-decile check has $s_1=0.12$ and an actual share of $0.16$.\footnote{No country holds more than 9 percent of the weakest direction. The bound of part (b) for that check, computed from $\Gamma$, $\lambda_2$, $q$ and $\beta_\tau$ alone, is $0.58$.} The valuable checks do involve hubs, but even the largest hub holds a small part of the weakest direction, so three members capture little of the value of repairing it.

\section{Bargaining over coherence}\label{sec:bargaining}

The formation wedge says that the members of a check that repairs a global inconsistency capture little of its value, so they do not build it. This section asks who can build such checks instead, and who can stop them. The answers depend on higher-order structure in three ways: on how a package of checks enters into force (Section~\ref{sec:package}), on whose links it must reach (Section~\ref{sec:table}), and on the shape of individual countries' partner graphs, which makes some of them indispensable (Section~\ref{sec:veto}), with consequences for how the gains can be divided (Section~\ref{sec:core}).

\subsection{Package agreements}\label{sec:package}

A single check has three members, and when it repairs a global inconsistency they capture little of its value. A package of checks negotiated together can enlist the countries on whose links that value falls. Whether it overcomes the wedge depends on how it enters into force. For a set $A$ of candidates, let $g_i(A)=L_i(T)-L_i(T\cup A)$ be country $i$'s fall in exposure when all of $A$ is added, $S(A)=\sum_ig_i(A)$, $M(A)$ the countries that are members of some check in $A$ (the \emph{participants}), and $s(A)=\sum_{i\in M(A)}g_i(A)/S(A)$ their share. Let $A_i\subseteq A$ be the checks of which $i$ is a member and
\[
\Delta_i(A)=g_i(A)-g_i(A\setminus A_i),
\]
the part of $i$'s gain that depends on its own checks. A package is offered to a set of signatories $N\supseteq M(A)$ with contributions $(p_i)_{i\in N}$, $\sum_ip_i=c|A|$; each signatory ratifies or not, and pays its contribution only if it ratifies and the package, or under partial entry at least its own checks, enters into force; under partial entry the contributions then cover the cost only when every check enters. Payoffs are the fall in exposure minus the contribution. Under \emph{all-or-nothing entry}, $A$ enters into force only if every signatory ratifies. Under \emph{partial entry}, each check enters into force when its own members ratify, so the package applies among those who ratify.

\begin{theorem}[Package agreements]\label{thm:package}
\econ Under all-or-nothing entry, ratifying a package is a weakly dominant strategy for every signatory whose contribution does not exceed its gain, and the only undominated choice when the contribution is strictly below the gain. It can therefore be financed by its participants alone exactly when the part of its value that falls on them covers its cost, and by all countries exactly when it raises total welfare; no outsider need contribute if the participants capture enough of the value. Under partial entry, a country can refuse, keep the benefit of the other checks and pay nothing. Each participant then pays at most the part of its gain that depends on its own checks, countries in no check pay nothing, and when the checks do not interact the package finances exactly what members would finance check by check.
\formal (a) Under all-or-nothing entry, if $p_i\le g_i(A)$ for every $i\in N$, ratifying is weakly dominant for every signatory; if $p_i>g_i(A)$ for some $i$, $A$ does not enter into force in any equilibrium in weakly undominated strategies. If $p_i<g_i(A)$ for every $i\in N$, every signatory ratifies in every equilibrium in weakly undominated strategies. Hence contributions summing to $c|A|$ that make ratification weakly dominant for all signatories exist if and only if $\sum_{i\in N}g_i(A)\ge c|A|$, and contributions under which ratification is the only undominated outcome exist if and only if the inequality is strict: for $N=M(A)$, the condition is $s(A)S(A)\ge c|A|$; for $N=V$, it is $S(A)\ge c|A|$.
(b) Under partial entry, $0\le\Delta_i(A)\le g_i(A)$, and ratification by all signatories is a Nash equilibrium if and only if $p_i\le\Delta_i(A)$ for every $i\in N$. Hence such contributions exist if and only if $\sum_{i\in M(A)}\Delta_i(A)\ge c|A|$, and $p_j\le0$ for every $j\in N\setminus M(A)$.
(c) If $b_\tau^\top H^{-1}b_{\tau'}=0$ for all distinct $\tau,\tau'\in A$, then $g_i(A)=\sum_{\tau\in A}g_i(\tau)$ and $\sum_{i}\Delta_i(A)=\sum_{\tau\in A}s(\tau)S(\tau)$.
\end{theorem}

The theorem turns the wedge into a question of institutional design. The value of checks that repair global inconsistencies falls on many countries, so no small group finances them; a package can enlarge the group that pays, but only if no participant can leave and keep the benefits. All-or-nothing entry makes every signatory pivotal, so each pays up to its full gain. Partial entry makes each participant pivotal only for its own checks; when the checks of a package substitute for one another, as checks acting on the same weak direction do, the part of a participant's gain that depends on its own checks is small, and the package is no easier to finance than its checks one by one. Part (c) gives the other extreme: when checks act on unrelated directions, partial entry is exactly check-by-check formation. This is the case of local gaps, in which a check's members bear most of its value anyway, so partial entry suffices there.

\subsection{Who must be at the table}\label{sec:table}

Which countries must take part? A check can be built only by its members, so a coalition $K\subseteq V$ can add only checks whose three links lie in $E(K)$, the links between its own members. Let $\mathcal C_K=\{z\in C:\ z_e=0\text{ for all }e\in E(K)\}$ be the directions of inconsistency that avoid those links, let $N_\mu$ be the number of eigenvalues of $H$ on $C$ below $\mu$, the dimension of the \emph{weak band}, and
\[
\Gamma(T;K)=\min_{z\in\mathcal C_K,\ \|z\|=1}z^\top Hz\qquad(\Gamma(T;K)=\infty\text{ if }\mathcal C_K=\{0\}).
\]

\begin{proposition}[Who must be at the table]\label{prop:table}
\econ A coalition can strengthen only the directions of inconsistency that run through links among its own members. However many checks it builds, it cannot raise the gap above the strength of the weakest direction that avoids those links. A coalition whose internal links miss the weak directions achieves nothing, whatever its size, and raising the gap requires at least as many internal links as there are directions to strengthen.
\formal (a) If every check in $A$ has all its members in $K$, then $\Gamma(T\cup A)\le\Gamma(T;K)$. (b) For every $\mu>0$, if $|E(K)|<N_\mu$ then $\Gamma(T;K)<\mu$. Hence a coalition that can raise the gap to $\mu$ has $|E(K)|\ge N_\mu$, so $|K|\ge\sqrt{2N_\mu}$ and $|K|\ge2N_\mu/d_{\max}$, where $d_{\max}$ is the largest number of links of any country.
\end{proposition}

Part (a) makes the identity of the participants matter, not their number; part (b) turns the number of weak directions into a minimum coalition size. In a local economy, $d_{\max}$ stays fixed while the weak directions multiply as the economy grows. On the grid of Proposition~\ref{S-prop:grid}, for instance, any coalition that raises the gap to $0.25$ must include at least about 3 percent of all countries at every size we compute.\footnote{There $d_{\max}=8$, and the number of directions weaker than $0.25$ stays between 11 and 14 percent of the number of countries for grids from $6\times6$ to $22\times22$.} Section~\ref{sec:bargevidence} shows that in the treaty network the bound of part (a) is close to what coalitions achieve.

\subsection{Indispensable countries}\label{sec:veto}

Proposition~\ref{prop:table} bounds what a given coalition can do. The converse question is which countries every successful coalition must include. The answer is largely local. For a country $x$ with at least two partners, let $\mathrm{Lk}(x)$ be its partner graph (Section~\ref{sec:arch}), whose vertices are all of $x$'s partners, including those with which it shares no check, with one edge $jk$ for each check $\{x,j,k\}$, $L_x$ its Laplacian, and $\lambda_2(L_x)$ the second-smallest eigenvalue, which is zero exactly when $\mathrm{Lk}(x)$ is disconnected. Let $L_G=d_0^*d_0$ be the Laplacian of the bilateral graph $G$ and $L_{G-x}$ that of $G$ without $x$, and let $\kappa_x$ be the largest value of $f^\top L_{G-x}^{+}f$ over unit vectors $f$ supported on the partners of $x$ with $\sum_jf_j=0$; it bounds the corresponding value for $L_G^+$, because adding $x$ and its links can only lower effective resistances. It is small when $x$'s partners are well connected to one another through the bilateral graph. For a set $A$ of checks none of which contains $x$, let $\Gamma_{-x}(A)$ be the gap of the checks of $T\cup A$ that do not contain $x$, computed on $G-x$: the coherence the other countries reach among themselves ($\Gamma_{-x}(A)=\infty$ if $G-x$ has no cycles). Let $\mu_x(A)$ be the smallest ratio of the energy those checks assign to a cycle of $G-x$ to the squared norm of the cycle on the edges of $\mathrm{Lk}(x)$: how strongly the others discipline the links among $x$'s partners that $x$'s own checks rely on ($\mu_x(A)=\infty$ if no cycle of $G-x$ uses those links; the bounds below are then read as limits). Finally, with $M_x$ and $M_x'$ the restrictions of $L_G^+$ and $L_{G-x}^+$ to $x$'s partners and $f$ ranging over vectors on the partners with $\sum_jf_j=0$, let
\[
\hat\lambda_x=\min_f\frac{f^\top L_xf}{f^\top(I-M_x)f},\qquad \nu_x=\min_f\frac{f^\top(I+M_x')L_x(I+M_x')f}{f^\top(I+M_x')f},
\]
the connectivity of the partner graph corrected for how the rest of the bilateral graph absorbs flows through $x$'s partners; both are close to $\lambda_2(L_x)$ when $\kappa_x$ is small.

\begin{theorem}[Indispensable countries]\label{thm:veto}
\econ However many checks are built without a given country, the coherence gap cannot exceed a bound set by that country's partner graph, corrected for how the rest of the bilateral graph absorbs flows through its partners. If its partners fall into groups that no check containing it joins, no coalition without it can make the economy coherent. Conversely, when the other countries are coherent among themselves and discipline the links among its partners strongly, the gap reached without it is at least a fixed fraction of the same partner-graph quantity, for instance at least half of it when $\gamma_{-x}\ge2\nu_x$ and $\mu_x\ge2$. A country's power to hold up coherence then comes from the local shape of its agreements, not from its size or the number of its agreements.
\formal Let no check in $A$ contain $x$, let $x$ have at least two partners and $G-x$ be connected, $\lambda=\lambda_2(L_x)$, $\gamma_{-x}=\Gamma_{-x}(A)$ and $\mu_x=\mu_x(A)$. Then $\nu_x\gamma_{-x}\mu_x/((1+\mu_x)\gamma_{-x}+\mu_x\nu_x)\le\Gamma(T\cup A)\le\hat\lambda_x$, and if also $\kappa_x<1$,
\begin{align*}
\frac{\lambda(1-\kappa_x)^2\gamma_{-x}}{(1+\gamma_{-x})(1+\kappa_x)+\lambda(1-\kappa_x)^2}&\ \le\ \frac{\mu_x\gamma_{-x}\nu_x}{(1+\mu_x)\gamma_{-x}+\mu_x\nu_x}\ \le\ \Gamma(T\cup A),\\
\Gamma(T\cup A)&\ \le\ \hat\lambda_x\ \le\ \frac{\lambda}{1-\kappa_x}.
\end{align*}
In particular, if $\mathrm{Lk}(x)$ is disconnected, $\Gamma(T\cup A)=0$ for every such $A$, and when $\mu_x$ is large and $\nu_x$ is small relative to $\gamma_{-x}$, the lower bound approaches $\nu_x$, which is close to $\hat\lambda_x$ when $\kappa_x$ is small.
\end{theorem}

Proposition~\ref{prop:local} goes from local to global: if every partner graph is well connected, the economy is coherent. Theorem~\ref{thm:veto} is the converse for a single country: one poorly connected partner graph caps the gap of every architecture that does not change it, and only checks containing that country can change it. The upper bound comes from a direction of inconsistency that rotates around $x$, weighted by the second eigenvector of its partner graph; no check without $x$ touches it. The lower bound splits any direction into a part around $x$, which only $x$'s checks discipline, and a part among the others, which their own checks discipline; the factor $\mu_x/(1+\mu_x)$ is the price of that split, and it is close to one when the others' checks bear heavily on the links among $x$'s partners. Unlike Garland's method, the argument needs no condition on the other partner graphs. The outer bounds are simpler to compute and need only $\lambda_2(L_x)$, $\kappa_x$ and $\gamma_{-x}$.

\subsection{The core of the coherence game}\label{sec:core}

Indispensability determines how the gains from coherence can be divided. Fix a target $\gamma>0$ and let a coalition $K$ \emph{win} if the checks among its members can raise the gap to $\gamma$: $\Gamma(T\cup A(K))\ge\gamma$, where $A(K)$ is the set of all candidates whose members lie in $K$. Winning is monotone, because adding checks never lowers the gap. We consider targets that all countries together can reach, $\gamma\le\bar\gamma=\Gamma(T\cup A(V))$. Reaching the target creates a surplus, normalized to one, that the countries can divide. When the gains are measured by the gap itself, coalition $K$ can secure $v(K)=\Gamma(T\cup A(K))-\Gamma(T)$, so that $v(V)=\bar\gamma-\Gamma(T)$. An allocation $y\ge0$ with $\sum_iy_i=1$ is in the \emph{core} if every winning coalition receives at least one, $\sum_{i\in K}y_i\ge1$, so that no winning coalition can do better on its own. Let $\mathrm{Vet}_\gamma$ be the set of countries whose exclusion makes the target unreachable, $\Gamma(T\cup A(V\setminus\{x\}))<\gamma$.

\begin{corollary}[The core of the coherence game]\label{thm:core}
\econ The core is empty unless some countries are indispensable for the target, and then it gives the whole surplus to the indispensable countries. Modest targets have no indispensable countries, so every division of the gains can be blocked. Ambitious targets create them; a poorly connected partner graph is enough to make a country indispensable, and in the data the indispensable countries are not the largest or most integrated.
\formal (a) The core is nonempty if and only if $\mathrm{Vet}_\gamma\neq\emptyset$, and then it is the set of allocations with $y_i=0$ for $i\notin\mathrm{Vet}_\gamma$. (b) Under the conditions of Theorem~\ref{thm:veto}, $x\in\mathrm{Vet}_\gamma$ whenever an upper bound of that theorem is below $\gamma$, and $x\notin\mathrm{Vet}_\gamma$ whenever a lower bound, with $A=A(V\setminus\{x\})$, is at least $\gamma$; a coalition $K$ loses whenever $\Gamma(T;K)<\gamma$. (c) $\mathrm{Vet}_\gamma$ is nondecreasing in $\gamma$; with $\underline\gamma=\min_x\Gamma(T\cup A(V\setminus\{x\}))$, the core is empty for $\gamma\le\underline\gamma$ and nonempty for $\underline\gamma<\gamma\le\bar\gamma$.
\end{corollary}

Part (a) is the classical characterization of the core of a simple game. The coherence game is stylized, but it isolates the force that any bargaining over a coherent architecture must face. Hold-up power over coherence belongs above all to the countries whose partners the existing agreements fail to join, and Section~\ref{sec:bargevidence} shows that in the treaty network they are not the countries one would expect.

The core is stark; Appendix~\ref{S-app:shapley} shows that the Shapley value spreads the gains more widely but rewards the same countries.

\textbf{Bargaining.} The core says which divisions are stable, not which one negotiations produce. Consider the protocol of \citet{baronferejohn1989}, applied to the coherence game with target $\gamma$. In each period a country, drawn with equal probability, proposes a coalition $W$ containing itself that can reach the target, so that every check of the package has all its members in $W$, together with a division of the surplus among the members of $W$. The package enters into force if every member of $W$ accepts; otherwise the next period begins, and payoffs are discounted by $\delta<1$ per period. We study stationary subgame perfect equilibria in which every proposal is accepted and responders accept offers equal to their continuation values.\footnote{Proposers may mix among equally cheap coalitions. \citet{banksduggan2000} establish existence of stationary equilibria for general versions of this protocol, and \citet{eraslanmclennan2013} show that stationary equilibrium payoffs are unique for arbitrary simple games.} Let $k=|\mathrm{Vet}_\gamma|$.

\begin{theorem}[Bargaining over coherence]\label{thm:bargain}
\econ In the equilibria studied, the indispensable countries receive equal shares, and as countries become patient they divide the whole surplus equally among themselves while every other country receives nothing. A country's bargaining power is therefore one over the number of indispensable countries if it is one of them, and zero otherwise, and Theorem~\ref{thm:veto} certifies membership from partner graphs. When the target is uncertain among the levels at which some country is indispensable, a country's expected share is the average of these reciprocals over the targets for which it is indispensable.
\formal Let $\mathrm{Vet}_\gamma\neq\emptyset$ and let $w_x(\delta)$ be the equilibrium payoff of $x$. (a) Every $x\in\mathrm{Vet}_\gamma$ receives $w_x(\delta)=(1-\delta\eta)/(\delta k+n(1-\delta))$, where $\eta$ is the smallest total equilibrium payoff of the dispensable countries that a winning coalition must include; $\eta$ is determined in equilibrium, and $\eta=0$ when $\mathrm{Vet}_\gamma$ is itself winning. (b) As $\delta\to1$, $w_x\to1/k$ for $x\in\mathrm{Vet}_\gamma$ and $w_x\to0$ otherwise. (c) Under the conditions of Theorem~\ref{thm:veto}, $x$ receives a positive limit share if $\hat\lambda_x<\gamma$ and none if its lower bound in that theorem, with $A=A(V\setminus\{x\})$, is at least $\gamma$. If the target is uniform on $(\underline\gamma,\bar\gamma]$, the range on which some country is indispensable, $x$'s expected limit share is $\Psi_x=\int_{\max\{c_x,\underline\gamma\}}^{\bar\gamma}d\gamma/|\mathrm{Vet}_\gamma|$ divided by $\bar\gamma-\underline\gamma$, with $c_x=\Gamma(T\cup A(V\setminus\{x\}))$.
\end{theorem}

The theorem makes the outcome of negotiations a function of indispensability, which Theorem~\ref{thm:veto} ties to the partner graphs. The countries with bargaining power are the indispensable ones, and how much power each has depends on how many others are indispensable for the same target: a country with a poorly connected partner graph gains when few others share its position and loses its share as the target rises and more countries become indispensable. Unlike the hold-up claims below, the protocol never leaves the gains unallocated: proposals buy the indispensable countries' consent at their continuation values, and as countries become patient competition drives the price of every dispensable country to zero; for $\delta<1$ it is at most $(n-1)(1-\delta)/\delta$.

Indispensability also limits how far bargaining can go. Let $m_x=v(V)-v(V\setminus\{x\})=\bar\gamma-\Gamma(T\cup A(V\setminus\{x\}))$ be country $x$'s marginal contribution to full coherence: what the others lose if $x$ withholds its consent. In bilateral negotiations in which a country can refuse and the others can proceed without it, each country can claim part of $m_x$.\footnote{With equal bargaining weights in Nash bargaining between the package and each country, the claim is $m_x/2$.} Call a division $y\ge0$ of $v(V)$ \emph{$\beta$-hold-up-proof} if $y_x\ge\beta m_x$ for every $x$.

\begin{proposition}[Hold-up]\label{prop:holdup}
\econ Coherence is a highly complementary good. The marginal contributions of the countries whose links carry the weak directions can add up to many times the total gain, so if each country can claim even a small fraction of what it adds, no division of the gains satisfies all claims. The upper bounds of Theorem~\ref{thm:veto} can certify this.
\formal (a) A $\beta$-hold-up-proof division exists if and only if $\beta\sum_xm_x\le v(V)$. (b) Set $\hat\lambda_x=\infty$ when $x$ has fewer than two partners or $G-x$ is disconnected. Then $m_x\ge(\bar\gamma-\hat\lambda_x)^+$ for every $x$, so no $\beta$-hold-up-proof division exists whenever $\beta\sum_x(\bar\gamma-\hat\lambda_x)^+>v(V)$.
\end{proposition}

The gap is set by the weakest direction, so when the weak directions run through many countries' links, each of them is nearly indispensable for full coherence, and a package that must buy every such country's consent at even a small fraction of its marginal value cannot be financed.

\subsection{A solution}\label{sec:solution}

Part (a) of Theorem~\ref{thm:package} clarifies why the participants are the natural signatories. Listing every country as a signatory would let the package internalize all of its value, but a country that can avoid being listed would free-ride; the participants cannot, because the checks require their consent. With Corollary~\ref{cor:expanders}, the results give a solution in four parts. Where to build: the weakest directions, through sparse, non-local checks (Theorem~\ref{thm:3}). Who must take part: the countries whose links those directions run through (Proposition~\ref{prop:table}), including every indispensable country (Theorem~\ref{thm:veto}), which can claim the gains from coherence (Corollary~\ref{thm:core}). Who pays: the participants of a package that enters into force only if all of them ratify, each in proportion to its own gain. What follows: once the core is coherent, sufficiently valuable additions can be left to pairwise incentives. Coordination is needed once. The package solves the free-rider problem behind the wedge, but not hold-up: Theorem~\ref{thm:package} takes the proposed contributions as given, and if the terms can be renegotiated, Corollary~\ref{thm:core} and Theorem~\ref{thm:bargain} say that indispensable countries claim the whole surplus. We do not model how the package and its terms are chosen, only which packages can be ratified and how the gains can be divided.

\section{Evidence from trade agreements}\label{sec:evidence}

We apply the theory to the network of trade agreements in DESTA 2.3 \citep{dur2014,desta2025}, restricted to the 80 economies of the OECD Inter-Country Input--Output list \citep{oecd2025icio}, treating common membership in a treaty as a check.\footnote{Appendix~\ref{S-app:tax} applies the same logic to bilateral tax treaties, where a cheaper route through a third country signals an inconsistent triple: among triples with three bilateral treaties, a detour is cheaper in 9 to 15 percent of cases.} Section~\ref{sec:repair} measures where the network is weak and designs a stronger architecture, Section~\ref{sec:incentives} computes who would capture the value of the missing checks and tests the formation wedge on accessions, and Section~\ref{sec:bargevidence} applies the results of Section~\ref{sec:bargaining}. Common membership is not an observed correction mechanism, so the diagnostics describe the institutional network in the terms the theory makes computable; the accession test is the one piece of evidence on behaviour.

\subsection{The network and its repair}\label{sec:repair}\label{sec:overlap}

Each treaty gives a link for every pair of its members and a checked triangle for every triple. In 2022 the 1,771 links form a connected and highly clustered network, with a clustering coefficient of 0.81: 17,651 of the 23,891 triangles whose three links exist (74 percent) reflect common membership in a treaty. Yet $\dim\H^1(X;\R)=76$, so the coherence gap is zero, and the number of uncovered cycles has stayed between 67 and 78 since 2016 (Table~\ref{S-tab:memfull}). Neither connectivity nor clustering detects these directions.

A \emph{candidate addition} is a triple of countries that already have all three bilateral treaties but no common treaty; there are 6,240.\footnote{Adding one means a plurilateral arrangement among existing treaty partners, so the design changes how agreements are organized, not who has a treaty with whom. The benchmark adds candidates at random after the same first stage.} A greedy procedure guided by the theory first makes every inconsistency visible, which takes 76 additions, the minimum possible, and then adds the candidates that act most strongly on the weakest direction (Appendix~\ref{S-app:data}).

\begin{table}[t]
\centering\small
\caption{Repairing the 2022 treaty network: coherence gap and worst-case amplification}\label{tab:repair}
\begin{tabular}{rcccc}
\toprule
Added three-party & \multicolumn{2}{c}{Gap $\Gamma$} & \multicolumn{2}{c}{Amplification $1/\Gamma$ (units of $1/\kappa$)}\\
\cmidrule(lr){2-3}\cmidrule(lr){4-5}
arrangements & Targeted & Random & Targeted & Random\\
\midrule
0 & 0 & 0 & unbounded & unbounded\\
76 & 0.0035 & 0.0035 & 286 & 286\\
101 & 0.126 & 0.024 & 7.9 & 41\\
126 & 0.291 & 0.035 & 3.4 & 28\\
176 & 0.591 & 0.083 & 1.7 & 12\\
6,240 (all candidates) & 0.973 & 0.973 & 1.03 & 1.03\\
\bottomrule
\end{tabular}

\smallskip
\parbox{0.92\linewidth}{\footnotesize\emph{Notes:} 80 economies, 1,771 bilateral links, 17,651 triples covered by a common treaty and 6,240 further triples whose three links exist. The first 76 additions are the first stage; later rows add the stated number of further checks in the second stage. Random is the mean of eight draws added to the same 76-check design. $\Gamma$ is the unweighted coherence gap of Section~\ref{sec:framework}. Amplification is the worst-case response of Theorem~\ref{thm:1} for unit capacity $\kappa=1$; for other $\kappa$ divide by $\kappa$.}
\end{table}

Removing every uncovered cycle is cheap but not enough: 76 arrangements lift the gap only to 0.0035. Fifty more, 0.3 percent of the checks in place, raise it to 0.29, eight times what the same number placed at random achieves, and by Proposition~\ref{S-prop:cost} cut the capacity each check needs by 99 percent. Placement matters more than quantity, and the right placement is non-local: two thirds of the 126 arrangements involve no EU member.\footnote{Their most frequent participants have many bilateral treaties but little common coverage among their partners (Appendix~\ref{S-app:data}).}

\subsection{Why the targeted checks are not built}\label{sec:incentives}

We start from the 2022 complex plus the 76 first-stage additions, so that $\Gamma=0.0035>0$ and exposure is finite.\footnote{By Remark~\ref{rem:invisible}, the first-stage checks are those members have a private reason to build.} For each of the remaining 6,164 candidates we compute $S(\tau)$ and $s(\tau)$ exactly (Proposition~\ref{prop:wedge}) and sort them by their first-order effect on $\Gamma$ (Table~\ref{S-tab:shares}).

Where candidates barely affect the weakest directions, members capture about half of the value. For the three deciles that matter most, they capture 15 to 16 percent, while the social value is three orders of magnitude larger.\footnote{The rank correlation between the gain in $\Gamma$ and the members' share is $-0.83$, and the 50 checks chosen by the design have a median share of 18 percent.}

\medskip\noindent\textbf{Evidence from accessions.}\label{sec:accession} The diagnostics above show that the valuable checks would be missing; they do not show that countries respond to the wedge. Accessions provide a test of that prediction. When a country joins a plurilateral agreement, every triple formed by the newcomer and two incumbents becomes checked; if the three links already exist, the accession adds checks in exactly the sense of Section~\ref{sec:formation}. The theory predicts that accessions follow the members' gain from these checks, not the gain to the system.

The risk set contains every agreement that admits new members and every economy not yet a member, from 1980 to 2020.\footnote{There are 40 such agreements; 270 of 70,023 country-agreement-years end in accession the next year.} For each, we compute exactly how much the accession's block of checks would lower every country's exposure in that year's network, and split the total into the members' gain (newcomer and incumbents), the outsiders' gain and the social value $S$ (Appendix~\ref{S-app:data}). The 101 accessions with a nonempty block of checks identify the estimates; they amount to 32 decisions.\footnote{A country joining the European Union accedes at once to 12 to 15 of its agreements, so we cluster by newcomer and year.}

\begin{table}[t]
\centering\small
\caption{Accessions follow the members' gain, not the system's}\label{tab:accession}
\begin{tabular}{lcccc}
\toprule
 & (1) & (2) & (3) & (4)\\
\midrule
log members' gain & 1.48 & 5.28 & 7.71 & \\
 & (0.40) & (1.21) & (3.34) & \\
log outsiders' gain & $-0.99$ & $-2.50$ & $-6.19$ & \\
 & (0.27) & (1.06) & (2.57) & \\
log social value $S$ & & & & 0.42\\
 & & & & (0.30)\\
log outsiders' share of $S$ & & & & $-1.00$\\
 & & & & (0.23)\\
Makes an inconsistency visible & 1.49 & 3.18 & --- & 1.40\\
 & (0.51) & (1.54) & & (0.49)\\
\addlinespace
Year and agreement fixed effects & yes & yes & yes & yes\\
Newcomer fixed effects & & yes & & \\
Sample & all & all & single accessions & all\\
Observations & 3,214 & 763 & 245 & 3,214\\
Accessions (decisions) & 101 (32) & 101 (32) & 15 (15) & 101 (32)\\
\bottomrule
\end{tabular}

\smallskip
\parbox{0.92\linewidth}{\footnotesize\emph{Notes:} Logit of accession at $t+1$ on characteristics of the checks the accession would add on existing links in the year-$t$ network. Gains are reductions in exposure~\eqref{eq:exposure}; members are the newcomer and the incumbents. All columns control for the log number of checks added, the share of incumbents to which the newcomer is already linked, the log number of incumbents and the newcomer's log degree. Observations in fixed-effect groups without accessions are dropped. Standard errors in parentheses, clustered by newcomer and year (515, 156, 211 and 515 clusters). In column (3) the visibility indicator is dropped because it is almost collinear with the outcome in 15 events.}
\end{table}

Accessions are more likely the more the members gain, and, holding that fixed, less likely the more outsiders gain (Table~\ref{tab:accession}). Column (4) is the most direct test of the formation wedge the data allow. It separates the two benchmarks of Proposition~\ref{prop:wedge}: a planner, or countries that internalized the value they create, would respond to the social value $S$ and not to how it is split, so the coefficient on the outsiders' share would be zero; under the wedge, members respond to the share they keep, so it is negative. Holding the social value fixed, a larger outsiders' share deters accession ($z=-4.3$), and no draw of the link-preserving null below reaches that value. Countries respond to the share of value they keep, not to the value they create.\footnote{\label{fn:triples}The gain that matters is the incumbents', which fits agreements whose incumbents decide on admission, and accessions that make an invisible inconsistency visible are more likely, as Remark~\ref{rem:invisible} predicts. The incumbents' gain has coefficient 3.11 (standard error 0.63), the newcomer's $-0.40$ (0.12). The same test across single triples is much weaker because private and social values are almost collinear there (Appendix~\ref{S-app:data}).}

The pattern survives gains measured two years earlier, disappears when country labels are permuted, and survives a null that keeps every bilateral link and redraws only which triangles are checked, so it is carried by which triples are checked, not by who is linked.\footnote{Gains at $t-2$ give coefficients of 1.05 (0.42) and $-0.73$ (0.31). Across 20 label permutations the $z$-statistics average $-0.2$ and $0.3$, none as extreme as the observed $3.7$ and $-3.7$. The higher-order component of the gains, observed minus the average under the link-preserving null, gives 1.36 (0.38) and $-0.82$ (0.27) (Appendix~\ref{S-app:data}).} But 100 of the 101 informative accessions are to agreements with EU incumbents.

The evidence rests on 32 decisions, almost all involving EU-centred agreements, and is an association: the placebo rules out explanations unrelated to network structure, but not an omitted structural factor such as remoteness from existing blocs. It is, however, the pattern the formation wedge predicts, and not the pattern that a view of agreements as tools for system-wide coherence would predict.\footnote{The model predicts that outsiders' gains are irrelevant, not harmful; the negative coefficient suggests a wedge stronger than the model implies, for instance because incumbents are reluctant to provide outsiders a service for which they could otherwise charge.}

\subsection{Bargaining over the treaty network}\label{sec:bargevidence}

Built as one package, the 50 targeted checks of Section~\ref{sec:repair} look different, and how the package enters into force decides whether it is built (Theorem~\ref{thm:package}). They involve 47 of the 80 economies. Together they cut total exposure by 79 percent, every economy gains, and the 47 participants capture 92 percent of the package's value, against a median of 18 percent for the members of each check taken alone. Under all-or-nothing entry, the participants can therefore finance the package among themselves, with no contribution from outsiders and no participant worse off, whenever its cost is below 92 percent of its value.\footnote{Ten economies, led by Iceland, Norway and Switzerland, would pay just over half.} Under partial entry, the parts of the participants' gains that depend on their own checks add up to only 14 percent of the package's value, because the checks act on the same weak directions and each participant's own checks add little once the others are in place.\footnote{The package is worth 12 percent of the sum of its checks' stand-alone values.} The rule of entry into force, not the size of the package, is what overcomes the wedge. The OECD's Multilateral Instrument for tax treaties applies only between parties that both list a treaty, a form of partial entry; it works there because those gaps are local, as the end of this section shows.

Who must be in the package? Proposition~\ref{prop:table} says that a coalition can only strengthen directions that run through its own links, and in the treaty network this decides the answer. If the 28 economies of the EU-28 added every candidate check among themselves, the gap would stay at 0.0035: the weak directions avoid the links between them, and part (a) caps what they can reach at 0.007. A coalition of 27 economies chosen greedily for its effect on the gap, mostly outside the EU and led by Iceland, Switzerland, China, Korea and Turkey, raises the gap to 0.33 by adding every check among its members, while five random coalitions of the same size reach between 0.005 and 0.051.\footnote{Part (a) caps the greedy coalition at 0.41.} Membership, not size, determines what a coalition can repair.

Theorem~\ref{thm:veto} identifies which economies every repairing coalition must include. If all 80 economies added every candidate check, the gap would rise from 0.0035 to $\bar\gamma=0.97$. Excluding a single economy leaves it at 0.97 for most of them, but a few are indispensable (Table~\ref{tab:veto}). Without Ukraine, no set of checks can raise the gap above 0.071; without Costa Rica, above 0.079. Their partner graphs are poorly connected: their partners belong to blocs that no agreement including them joins, so only checks containing them can strengthen the directions that run through them. The bounds of Theorem~\ref{thm:veto} bracket the actual cap closely for the economies that matter.\footnote{The upper bound is within 10 percent of it for nine of the ten economies in Table~\ref{tab:veto} (48 percent for Kazakhstan), and the lower bound, where it applies, is between 68 and 84 percent of it.} Together they classify the economies without computing any coalition's gap: they certify every indispensable economy for targets of 0.1, 0.29 and 0.5, and 59 of the 70 others for a target of 0.29. Germany and France leave the gap at 0.97 when excluded, and across all 80 economies the cap is uncorrelated with the number of partners.\footnote{Germany's and France's partner graphs have $\lambda_2=25$, and the lower bound alone certifies that they are not indispensable for any target up to 0.93. The correlation between the cap and the number of partners is $-0.003$.}

\begin{table}[t]
\centering\small
\caption{Indispensable economies in the 2022 treaty network}\label{tab:veto}
\resizebox{\linewidth}{!}{\setlength{\tabcolsep}{4pt}\begin{tabular}{lrccccc}
\toprule
& & Partner-graph & \multicolumn{2}{c}{Bounds of Theorem~\ref{thm:veto}} & Gap without & Shapley\\
\cmidrule(lr){4-5}
Economy & Partners & $\lambda_2(L_x)$ & lower & upper & this economy & share (\%)\\
\midrule
Ukraine & 36 & 0.070 & 0.055 & 0.072 & 0.071 & 9.2\\
Costa Rica & 39 & 0.079 & 0.067 & 0.081 & 0.079 & 6.8\\
Hong Kong & 17 & 0.140 & 0.113 & 0.146 & 0.135 & 3.8\\
Jordan & 41 & 0.170 & 0.132 & 0.176 & 0.171 & 4.7\\
Kazakhstan & 33 & 0.243 & 0 & 0.255 & 0.173 & 3.7\\
Japan & 49 & 0.179 & 0 & 0.189 & 0.180 & 4.3\\
Chile & 63 & 0.190 & 0.141 & 0.199 & 0.182 & 4.1\\
Canada & 47 & 0.195 & 0.132 & 0.200 & 0.194 & 4.3\\
United States & 15 & 0.221 & 0.174 & 0.227 & 0.219 & 5.2\\
South Africa & 43 & 0.262 & 0.202 & 0.268 & 0.262 & 4.3\\
\addlinespace
Germany & 56 & 25.0 & 0.934 & 25.4 & 0.971 & 0.0\\
\bottomrule
\end{tabular}}

\smallskip
\parbox{0.92\linewidth}{\footnotesize\emph{Notes:} Starting point as in Table~\ref{S-tab:shares}. The last column is the gap when every candidate check that does not contain the economy is added; with all candidates the gap is 0.97. The bounds are the inner bounds of Theorem~\ref{thm:veto} with $A$ the candidates not containing the economy; the lower bound is zero for Kazakhstan and Japan because the other economies' checks leave an inconsistency among themselves undetected ($\gamma_{-x}=0$). The Shapley share is $\phi_x/v(V)$ for the game of Proposition~\ref{S-prop:shapley}, estimated from 120 random orders (standard errors 1 to 2 points for the economies shown). The ten economies shown are those with the lowest gap without them.}
\end{table}

By Corollary~\ref{thm:core}, the core of the coherence game is empty for every target up to $\underline\gamma=0.071$: no economy is indispensable, and every division of the gains can be blocked. Above it the core is nonempty and gives the whole surplus to the indispensable economies: Ukraine alone for targets up to 0.079, the ten economies of Table~\ref{tab:veto} for a target of 0.29, the gap the repair design reaches, and 22 economies for a target of 0.5. The economies with hold-up power over coherence are mid-sized bridges between blocs, not the largest or most integrated members of the network. The Shapley value softens the core without changing its message: the ten economies of Table~\ref{tab:veto} receive half of it (Appendix~\ref{S-app:shapley}).

Theorem~\ref{thm:bargain} turns these sets into bargaining power. For a target of 0.29, the ten economies of Table~\ref{tab:veto} would each receive a tenth of the surplus as countries become patient, and all others nothing. With the target uniform over the range on which some economy is indispensable, from 0.071 to 0.97, Ukraine's expected share is 10.6 percent and Costa Rica's 9.6 percent, and the ten economies together receive 60 percent. These equilibrium shares are close to the Shapley value, with a correlation of 0.95 across the 80 economies, although the two are derived from different principles.

Hold-up is severe (Proposition~\ref{prop:holdup}). Excluding any one of 26 economies would lower the gap reachable with all candidate checks by more than 0.1, and the marginal contributions $m_x$ sum to 17.0, against a total gain of $v(V)=0.97$. No division of the gains can give every economy more than 5.7 percent of its marginal contribution; the upper bounds of Theorem~\ref{thm:veto} alone certify that no division can give every economy more than 6.0 percent. With equal weights in bilateral Nash bargaining, the claims would exceed the gains ninefold. All ten economies of Table~\ref{tab:veto} are participants in the targeted package and gain from it, and the more indispensable an economy is, the more it tends to gain, so proportional terms usually reward hold-up power with a large stake.\footnote{Together the ten receive 28 percent of the package's value and, under contributions proportional to gains, pay 30 percent of its cost. The rank correlation between indispensability and share of the value is 0.72 across the 80 economies.} Kazakhstan is the exception: nearly as indispensable as Japan, it receives 0.3 percent of the value, against a Shapley share of 3.7 percent. Securing its participation is cheap at proportional terms, but its power to block the package is out of proportion to its stake, which is where hold-up is most likely.

Theorem~\ref{thm:package}(c) also predicts when partial entry suffices. Among bilateral tax treaties under the OECD's Multilateral Instrument, each unchecked inconsistency sits on a single treaty, its two partners bear about 98 percent of the exposure it causes, and 89 percent of 3,207 listing decisions cover the treaty, as the model predicts for local gaps (Appendix~\ref{S-app:mli}).

\section{Conclusion}

Bilateral links determine who can exchange. Higher-order structure determines whether the resulting arrangements can be reconciled, and, as this paper shows, who can reconcile them. The coherence gap governs how long inconsistencies persist, local clusters of checks lose it as economies grow and high-dimensional expanders keep it. The checks that would restore it have diffuse benefits, so their members do not build them, although once coherence is reached sufficiently valuable further checks pay for themselves.

The same structure governs the alternatives. A package of checks that substitute for one another overcomes the wedge only if it enters into force all-or-nothing; a coalition can repair only the directions of inconsistency that run through its own links; and a country whose partners no common agreement joins is indispensable, so that in the core, and in equilibrium bargaining, such countries take the gains from coherence. Because coherence is so complementary, bilateral hold-up claims cannot all be met. In the trade-agreement network, the EU cannot repair the system on its own, a coalition of bridging economies can, and Ukraine and Costa Rica can hold it up. Multilateral institutions therefore have a specific role: not to replace bilateral agreements, but to organize the checks among them whose benefits fall mostly on outsiders, through packages that enter into force only as a whole and that include the indispensable countries.

The results open several questions. The equilibrium analysis covers negotiations that end without delay; a theory in which hold-up produces delay or partial packages, and in which coalitions form over time, is the natural next step, and the minimal coalitions needed to fill a given inconsistency connect it to isoperimetric questions on simplicial complexes. Empirically, the persistence predictions can be tested where compatibility obligations are measured clause by clause, and the identification of indispensable countries can be compared with observed negotiating power.

\appendix
\section{Definitions and proofs}\label{app:proofs}

\textbf{Notation.} Let $d_0:\R^V\to\R^E$ be the signed participant-to-link difference operator, mapping benchmark values to bilateral adjustments, and $d_1:\R^E\to\R^T$ the map taking bilateral adjustments to the left-hand sides of the three-party checks \eqref{eq:2}, so that $d_1d_0=0$. Let $C=(\operatorname{im}d_0)^\perp$ be the space of adjustments orthogonal to benchmark differences; its dimension is $r=m-n+1$. Let $D=d_1|_C$ and $H=D^*D$, a symmetric nonnegative operator. The cohomology groups used below are $\H^1(X;\R)=\ker d_1/\operatorname{im}d_0$ (locally consistent adjustments modulo benchmark differences) and, for a two-dimensional complex, $\H^2(X;\R)=\R^T/\operatorname{im}d_1$ (check patterns not generated by any adjustment); over $\F_2$ the same definitions apply with arithmetic mod 2. The coherence gap is $\Gamma(X)=\lambda_{\min}(H)$. A zero gap means a nontrivial inconsistency direction escapes every check; a positive gap means all such directions are detected quantitatively. Here ``high-dimensional'' refers to restrictions involving groups of participants, not to a large number of commodities or state variables.

\subsection*{Proofs}

This appendix proves the formation results of Section~\ref{sec:formation}. Proofs of the results of Sections~\ref{sec:framework} to~\ref{sec:arch} are in Appendix~\ref{S-app:additional}.

\medskip
\noindent\emph{Proof of Proposition~\ref{prop:wedge}.} The boundary vector $b$ of a candidate lies in $C$, because $d_1d_0=0$. Adding the check replaces $H$ by $H+bb^\top$ on $C$. By the Sherman--Morrison formula, $(H+bb^\top)^{-1}=H^{-1}-uu^\top/(1+q)$ with $u=H^{-1}b$ and $q=b^\top u>0$. The stationary covariance is $\Sigma=H^{-1}$ under the normalization, so the variance on link $e$ falls by $u_e^2/(1+q)\ge0$, and country $i$'s exposure falls by $\sum_{e\ni i}u_e^2/(1+q)$. Summing over the three members counts link $e$ exactly $\nu_e(\tau)$ times, and summing over all countries counts it twice, which gives $S(\tau)$ and $s(\tau)$; $s\le1$ because $\nu_e\le2$. The members' joint gain is $s(\tau)S(\tau)$ and the planner's objective changes by $c-S(\tau)$, which gives the two conditions and the wedge.

\noindent\emph{Proof of Theorem~\ref{thm:general}.} Write $L_i(z)=\sum_{e\ni i}z_e^2$ for $z\in\R^E$, so that by Proposition~\ref{prop:wedge} $s(\tau)S(\tau)=\sum_{i\in\tau}L_i(u)/(1+q)$ and $s(\tau)=\sum_{i\in\tau}L_i(u)/(2\|u\|^2)$.

(a) Let $\pi_e$ be the orthogonal projection onto $C$ of the unit vector of link $e$, so that $u_e=\pi_e^\top H^{-1}b$ and $\Sigma_{ee}=\pi_e^\top H^{-1}\pi_e$. By the Cauchy--Schwarz inequality for the inner product defined by $H^{-1}$ on $C$, $u_e^2\le\Sigma_{ee}\,q$; summing over the members' links with multiplicities $\nu_e(\tau)$ gives $s(\tau)S(\tau)\le\frac{q}{1+q}\sum_{i\in\tau}L_i(T)$. Next, $\|u\|^2=\sum_j\lambda_j^{-2}(\varphi_j^\top b)^2\ge\beta_\tau^2/\Gamma^2$ and $q\le\|b\|^2/\Gamma=3/\Gamma$, so $S(\tau)\ge2\beta_\tau^2/(\Gamma(\Gamma+3))$. Each link of $\tau$ has $\nu_e=2$, so $\sum_{i\in\tau}L_i(u)\ge2\sum_{e\in\tau}u_e^2$, and since $q=\sum_{e\in\tau}b_eu_e$ with $b_e=\pm1$, the Cauchy--Schwarz inequality gives $q^2\le3\sum_{e\in\tau}u_e^2$. Hence $s(\tau)S(\tau)\ge2q^2/(3(1+q))$. Finally $\|u\|^2=b^\top H^{-2}b\le q/\Gamma$, so $s(\tau)\ge(2q^2/3)/(2q/\Gamma)=\Gamma q/3$.

(b) Write $u=a\varphi_1+r$ with $r\perp\varphi_1$, so that $a=\varphi_1^\top u=\beta_\tau/\Gamma$ (taking the sign of $\varphi_1$ so that $\varphi_1^\top b=\beta_\tau$) and $\varepsilon=\|r\|/\|u\|$, $|a|/\|u\|=\sqrt{1-\varepsilon^2}$. Let $N_\tau(z)=\big(\sum_e\nu_e(\tau)z_e^2\big)^{1/2}$, a seminorm with $N_\tau(z)^2\le2\|z\|^2$ because $\nu_e\le2$, and with $s(\tau)=N_\tau(u)^2/(2\|u\|^2)$ and $N_\tau(\varphi_1)^2=2s_1(\tau)$. By Minkowski's inequality, $|a|\,N_\tau(\varphi_1)-N_\tau(r)\le N_\tau(u)\le|a|\,N_\tau(\varphi_1)+N_\tau(r)$; dividing by $\sqrt2\|u\|$ gives $\sqrt{1-\varepsilon^2}\sqrt{s_1}-\varepsilon\le\sqrt{s(\tau)}\le\sqrt{s_1}+\varepsilon$. For the bound on $\varepsilon$, expand in the eigenvectors $\varphi_j$ of $H$, with $\beta_j=\varphi_j^\top b$: $\|r\|^2=\sum_{j\ge2}\beta_j^2/\lambda_j^2\le\lambda_2^{-1}\sum_{j\ge2}\beta_j^2/\lambda_j=\lambda_2^{-1}\big(q-\beta_\tau^2/\Gamma\big)$, while $\|u\|^2\ge a^2=\beta_\tau^2/\Gamma^2$. Dividing gives $\varepsilon^2\le(\Gamma/\lambda_2)(\Gamma q/\beta_\tau^2-1)$. \hfill$\square$

\noindent\emph{Proof of Corollary~\ref{cor:expanders}.} The matrix $DD^*$ on $\R^T$ has entries $b_\tau^\top b_{\tau'}$: 3 on the diagonal, $\pm1$ when $\tau$ and $\tau'$ share a link, and 0 otherwise. Each of the three links of $\tau$ lies in at most $\bar t-1$ other checks, so by Gershgorin's theorem every eigenvalue of $DD^*$ is at most $3+3(\bar t-1)=3\bar t$. The nonzero eigenvalues of $DD^*$ are those of $D^*D$, and $H$ is the restriction of $D^*D$ to the invariant subspace $C$, so $\lambda_{\max}(H)\le3\bar t$. For a candidate with boundary $b\in C$, $q=b^\top H^{-1}b\ge\|b\|^2/\lambda_{\max}(H)\ge1/\bar t$, and Theorem~\ref{thm:general}(a) gives $s(\tau)\ge\Gamma q/3\ge\gamma_0/(3\bar t)$. Hence $s(\tau)S(\tau)\ge c$ whenever $S(\tau)\ge3\bar tc/\gamma_0$. In $Y_n$ every link lies in exactly 9 checks, since $3|T(Y_n)|/|E(Y_n)|=3\cdot219/73=9$ and the quotients are regular, and Theorem~\ref{thm:3} gives $\gamma_0=9-4\sqrt2$, so $s(\tau)\ge(9-4\sqrt2)/27>0.12$. \hfill$\square$

\noindent\emph{Proof of Theorem~\ref{thm:package}.} Adding a set of checks adds a positive semidefinite matrix to $H$, so $H^{-1}$ falls in the positive semidefinite order on $C$; every diagonal entry of $\Sigma$ weakly falls, and so does every $L_i$. Hence $g_i$ is nondecreasing in the set of checks added, which gives $g_i(A)\ge0$ and $0\le\Delta_i(A)\le g_i(A)$.

(a) Not ratifying gives $i$ a payoff of 0 whatever the others do. Ratifying gives $g_i(A)-p_i$ if every other signatory ratifies and 0 otherwise. If $p_i\le g_i(A)$, ratifying is therefore weakly dominant; if $p_i<g_i(A)$, it is strictly better when every other signatory ratifies, so not ratifying is weakly dominated and every signatory ratifies in every equilibrium in weakly undominated strategies. If $p_i=g_i(A)$, $i$ is indifferent in every profile, so not ratifying is also undominated. If $p_i>g_i(A)$, ratifying is weakly dominated by not ratifying, so in weakly undominated strategies $i$ does not ratify and $A$ does not enter into force. If $\sum_{i\in N}g_i(A)\ge c|A|$, the contributions $p_i=c|A|\,g_i(A)/\sum_{j\in N}g_j(A)$ sum to $c|A|$ and satisfy $p_i\le g_i(A)$; conversely, $p_i\le g_i(A)$ for all $i$ and $\sum_ip_i=c|A|$ imply $\sum_{i\in N}g_i(A)\ge c|A|$. If the inequality is strict, subtracting $(\sum_{j\in N}g_j(A)-c|A|)/|N|$ from each $g_i(A)$ gives contributions with $p_i<g_i(A)$ for all $i$ that sum to $c|A|$; conversely, $p_i<g_i(A)$ for all $i$ forces strict inequality. The two special cases follow from the definitions of $s(A)$ and $S(A)$.

(b) Suppose every other signatory ratifies. If $i$ ratifies, all of $A$ enters into force and $i$ receives $g_i(A)-p_i$. If $i$ does not, the checks in $A_i$ do not enter into force, every other check does, because its members ratify, and $i$ pays nothing, so $i$ receives $g_i(A\setminus A_i)$. Ratifying is a best response if and only if $p_i\le\Delta_i(A)$. For $j\notin M(A)$, $A_j=\emptyset$ and $\Delta_j(A)=0$. Contributions with $p_i\le\Delta_i(A)$ and $\sum_ip_i=c|A|$ exist if and only if $\sum_{i\in N}\Delta_i(A)=\sum_{i\in M(A)}\Delta_i(A)\ge c|A|$.

(c) Let $B$ have the boundary vectors of $A$ as columns, $U=H^{-1}B$ and $G=B^\top H^{-1}B$. The condition says that $G$ is diagonal, with entries $q_\tau$, so by the Woodbury formula the fall in $\Sigma$ is $U(I+G)^{-1}U^\top=\sum_{\tau\in A}u_\tau u_\tau^\top/(1+q_\tau)$, the sum of the falls caused by each check alone (Proposition~\ref{prop:wedge}). Hence $g_i(A)=\sum_{\tau\in A}g_i(\tau)$, and the same holds for every subset of $A$, so $\Delta_i(A)=\sum_{\tau\in A_i}g_i(\tau)$. Summing over $i$ and exchanging the sums, $\sum_i\Delta_i(A)=\sum_{\tau\in A}\sum_{i\in\tau}g_i(\tau)=\sum_{\tau\in A}s(\tau)S(\tau)$. \hfill$\square$

\noindent\emph{Proof of Proposition~\ref{prop:table}.} (a) A check $\tau$ whose members lie in $K$ has its three links in $E(K)$, so its boundary vector $b_\tau$ vanishes off $E(K)$ and $b_\tau^\top z=0$ for every $z\in\mathcal C_K$. Hence $z^\top H_{T\cup A}z=z^\top Hz+\sum_{\tau\in A}(b_\tau^\top z)^2=z^\top Hz$ on $\mathcal C_K\subseteq C$, and $\Gamma(T\cup A)=\min_{z\in C,\|z\|=1}z^\top H_{T\cup A}z\le\Gamma(T;K)$. (b) The span $W_\mu$ of the eigenvectors of $H$ in $C$ with eigenvalues below $\mu$ has dimension $N_\mu$. The conditions $z_e=0$, $e\in E(K)$, are $|E(K)|<N_\mu$ linear restrictions, so $W_\mu$ contains a unit vector $z\in\mathcal C_K$, and $z^\top Hz<\mu$. For the size bounds, $|E(K)|\le|K|(|K|-1)/2<|K|^2/2$ and $|E(K)|\le|K|d_{\max}/2$. \hfill$\square$

\noindent\emph{Proof of Theorem~\ref{thm:veto}.} Orient every link at $x$ away from $x$. For $f\in\R^{N(x)}$ with $\sum_jf_j=0$, where $N(x)$ is the set of $x$'s partners, let $z\in\R^E$ equal $f_j$ on the link from $x$ to $j$ and zero elsewhere. A check contains a link at $x$ only if it contains $x$. For a check $\{x,j,k\}$, $b^\top z=\pm(f_j-f_k)$, and every other check of $T\cup A$, including all of $A$, has $b^\top z=0$. Write $Q=D^\top D$ for the matrix of the checks of $T\cup A$ on all of $\R^E$, so that $H_{T\cup A}$ is its restriction to $C$. Hence $z^\top Qz=\sum_{jk\in\mathrm{Lk}(x)}(f_j-f_k)^2=f^\top L_xf$. Let $P$ be the orthogonal projection onto $C=(\operatorname{im}d_0)^\perp$. Since $Dd_0=0$, $Qd_0=0$ and $(Pz)^\top H_{T\cup A}(Pz)=z^\top Qz$. The removed part is $(I-P)z=d_0L_G^+d_0^*z$, and $d_0^*z$ equals $f$ on $N(x)$ and $-\sum_jf_j=0$ at $x$, so $\|(I-P)z\|^2=f^\top M_xf$ and $\|Pz\|^2=f^\top(I-M_x)f$. This is positive for $f\neq0$: if $Pz=0$, $z=d_0\phi$ vanishes on the links of $G-x$, so $\phi$ is constant on the connected graph $G-x$, $f_j=\phi_j-\phi_x$ is constant, and $\sum_jf_j=0$ forces $f=0$. Moreover $f^\top M_xf\le f^\top M_x'f\le\kappa_x\|f\|^2$, because effective resistances between $x$'s partners are lower in $G$ than in $G-x$ (Rayleigh monotonicity). Hence $\Gamma(T\cup A)\le f^\top L_xf/f^\top(I-M_x)f$ for every such $f$, and minimizing gives $\Gamma(T\cup A)\le\hat\lambda_x$. Taking $f$ to be an eigenvector of $L_x$ for $\lambda_2(L_x)$, which has mean zero, gives $\hat\lambda_x\le\lambda_2(L_x)/(1-\kappa_x)$.

For the lower bound, write $\lambda=\lambda_2(L_x)$, $\kappa=\kappa_x$, $\nu=\nu_x$, $\gamma=\gamma_{-x}$, $\mu=\mu_x$, and let $z\in C$. Let $f\in\R^{N(x)}$ be its values on the links from $x$ to its partners and $z_R$ its restriction to the links of $G-x$. Since $d_0^*z=0$, $\sum_jf_j=0$ and, at each partner $j$, the net flow of $z_R$ out of $j$ equals $f_j$. Decompose $z_R=w+g$ with $w$ in the cycle space of $G-x$ and $g=d_0'\psi$ a gradient on $G-x$, where $d_0'$ is the coboundary of $G-x$; then $L_{G-x}\psi=-f$ (extending $f$ by zero), so $\psi=-L_{G-x}^+f$ and $\|g\|^2=f^\top M_x'f$. Checks not containing $x$ annihilate gradients and see only $z_R$, so their contribution to $z^\top H_{T\cup A}z$ is $w^\top H_{-x}w$, where $H_{-x}$ is the matrix of those checks on $G-x$; by the definitions of $\gamma$ and $\mu$, $w^\top H_{-x}w\ge\max\{\gamma\|w\|^2,\mu\|w_{\mathrm{Lk}}\|^2\}$, where $w_{\mathrm{Lk}}$ is the restriction of $w$ to the edges of $\mathrm{Lk}(x)$. A check $\{x,j,k\}$ contributes $(F_j-F_k+w_{jk})^2$ with $F=f-\psi|_{N(x)}=(I+M_x')f$, because $g_{jk}=\psi_k-\psi_j$; each link $jk$ lies in at most one check with $x$, so these contributions sum to $\|\delta_xF+w_{\mathrm{Lk}}\|^2$, where $\delta_xF$ has entries $F_j-F_k$ over the edges of $\mathrm{Lk}(x)$ and $\|\delta_xF\|^2=f^\top(I+M_x')L_x(I+M_x')f=:q$. Since $w$ and $g$ are orthogonal, $\|z\|^2=\|f\|^2+\|g\|^2+\|w\|^2=p+\|w\|^2$ with $p:=f^\top(I+M_x')f$. With $a=\sqrt q$, $c=\|w_{\mathrm{Lk}}\|$ and $b=\|w\|$, the triangle inequality gives
\begin{align*}
z^\top H_{T\cup A}z&\ \ge\ \max\{\mu c^2,\gamma b^2\}\ \text{and}\ \ge\ \mu c^2+\big((a-c)^+\big)^2\ \ge\ \frac{\mu a^2}{1+\mu}=:K,\\
\|z\|^2&\ =\ p+b^2 .
\end{align*}
If $\gamma b^2\le K$, the ratio $z^\top H_{T\cup A}z/\|z\|^2$ is at least $K/(p+K/\gamma)$; otherwise it is at least $\gamma b^2/(p+b^2)$, which increases in $b$ and so is again at least $K/(p+K/\gamma)$. Since $q\ge\nu p$, $K/(p+K/\gamma)=\big(\tfrac{\mu}{1+\mu}\big)\big/\big(\tfrac pq+\tfrac{\mu}{(1+\mu)\gamma}\big)\ge\mu\gamma\nu/((1+\mu)\gamma+\mu\nu)$. If $f=0$, then $z=w$ and the ratio is at least $\gamma$, which is larger. This proves the inner lower bound. For the outer one, $\mu\ge\gamma$ because $\|w_{\mathrm{Lk}}\|\le\|w\|$, and the bound increases in $\mu$ and in $\nu$; moreover $\|P_\perp F\|\ge\|f\|-\|P_\perp M_x'f\|\ge(1-\kappa)\|f\|$, where $P_\perp$ removes the mean on $N(x)$, so $q\ge\lambda(1-\kappa)^2\|f\|^2$ and $p\le(1+\kappa)\|f\|^2$, giving $\nu\ge\lambda(1-\kappa)^2/(1+\kappa)$. Substituting $\mu=\gamma$ and this value of $\nu$ gives the outer lower bound. \hfill$\square$

\medskip
\noindent\emph{Proof of Corollary~\ref{thm:core}.} Adding checks never lowers the gap, so $A(K)\subseteq A(K')$ for $K\subseteq K'$ makes winning monotone. (a) Let $y$ be in the core. For $i\notin\mathrm{Vet}_\gamma$, the coalition $V\setminus\{i\}$ wins, so $\sum_{j\ne i}y_j\ge1$ and $y_i=0$. If $\mathrm{Vet}_\gamma=\emptyset$ this gives $y=0$, contradicting $\sum_iy_i=1$, so the core is empty. If $\mathrm{Vet}_\gamma\neq\emptyset$, every winning coalition contains $\mathrm{Vet}_\gamma$, since a coalition missing $x\in\mathrm{Vet}_\gamma$ is contained in $V\setminus\{x\}$, which loses; so any allocation with $y_i=0$ off $\mathrm{Vet}_\gamma$ gives every winning coalition the whole surplus and is in the core. (b) Apply Theorem~\ref{thm:veto} with $A=A(V\setminus\{x\})$, and Proposition~\ref{prop:table}(a) with $A=A(K)$. (c) $\mathrm{Vet}_\gamma=\{x:\Gamma(T\cup A(V\setminus\{x\}))<\gamma\}$ grows with $\gamma$ and is empty exactly when $\gamma\le\underline\gamma$; apply (a). \hfill$\square$

\medskip
\noindent\emph{Proof of Proposition~\ref{prop:holdup}.} (a) If $\beta\sum_xm_x\le v(V)$, the division $y_x=\beta m_x+\big(v(V)-\beta\sum_jm_j\big)/n$ is nonnegative, sums to $v(V)$ and satisfies every claim. Conversely, summing $y_x\ge\beta m_x$ over $x$ gives $v(V)\ge\beta\sum_xm_x$. (b) Theorem~\ref{thm:veto} with $A=A(V\setminus\{x\})$ gives $\Gamma(T\cup A(V\setminus\{x\}))\le\hat\lambda_x$, so $m_x\ge\bar\gamma-\hat\lambda_x$; since $m_x\ge0$ by monotonicity, $m_x\ge(\bar\gamma-\hat\lambda_x)^+$. Apply (a). \hfill$\square$

\medskip
\noindent\emph{Proof of Theorem~\ref{thm:bargain}.} In a stationary equilibrium in which every proposal is accepted, the payoffs satisfy $\sum_xw_x=1$, a responder $j$ accepts an offer of at least $\delta w_j$, and a proposer $i$ offers exactly $\delta w_j$ to every other member of a cheapest winning coalition containing it, keeping $1-\delta C_i$ with $C_i=\min\{\sum_{j\in W\setminus\{i\}}w_j:\ W\ni i\text{ winning}\}$. Hence $w_i=\frac1n(1-\delta C_i)+\delta\rho_iw_i$, where $\rho_i$ is the unconditional probability that another country proposes and includes $i$, allowing proposers to mix among equally cheap coalitions. Every winning coalition contains $\mathrm{Vet}_\gamma$, so for $v\in\mathrm{Vet}_\gamma$, $\rho_v=(n-1)/n$ and $C_v=S-w_v+\eta$, with $S=\sum_{u\in\mathrm{Vet}_\gamma}w_u$ and $\eta=\min\{\sum_{j\in W\setminus\mathrm{Vet}_\gamma}w_j:\ W\text{ winning}\}$, which does not depend on $v$. Substituting gives $w_v\,n(1-\delta)=1-\delta(S+\eta)$, the same for every $v$, so $S=kw_v$ and (a) follows by solving. For (b), the veto equation gives $w_v(n-\delta(n-1))=1-\delta C_v$. For $j\notin\mathrm{Vet}_\gamma$, $V\setminus\{j\}$ wins and contains $v$, so $C_v\le1-w_v-w_j$ and $w_v(n-\delta(n-1))\ge\delta w_v+\delta w_j+1-\delta$, that is, $w_vn(1-\delta)\ge\delta w_j+1-\delta$. As $\delta\to1$ the left side vanishes, because $w_v\le1$, so $w_j\to0$; then $\eta\to0$, and (a) gives $w_v\to1/k$. (c) follows from (b), Theorem~\ref{thm:veto} and the definition of $\mathrm{Vet}_\gamma$, integrating over $\gamma$. \hfill$\square$

\clearpage
\setcounter{section}{0}\setcounter{proposition}{0}\setcounter{theorem}{0}\setcounter{lemma}{0}
\setcounter{remark}{0}\setcounter{corollary}{0}\setcounter{equation}{0}\setcounter{table}{0}
\renewcommand{\thesection}{S\arabic{section}}\renewcommand{\theHsection}{S\arabic{section}}
\renewcommand{\theproposition}{S\arabic{proposition}}\renewcommand{\theHproposition}{S\arabic{proposition}}
\renewcommand{\theremark}{S\arabic{remark}}\renewcommand{\theHremark}{S\arabic{remark}}
\renewcommand{\thetheorem}{S\arabic{theorem}}\renewcommand{\theHtheorem}{S\arabic{theorem}}
\renewcommand{\thelemma}{S\arabic{lemma}}\renewcommand{\theHlemma}{S\arabic{lemma}}
\renewcommand{\thecorollary}{S\arabic{corollary}}\renewcommand{\theHcorollary}{S\arabic{corollary}}
\renewcommand{\theequation}{S\arabic{equation}}\renewcommand{\theHequation}{S\arabic{equation}}
\renewcommand{\thetable}{S\arabic{table}}\renewcommand{\theHtable}{S\arabic{table}}
\begin{center}{\Large\textbf{Supplemental Appendix}}\end{center}
\noindent References to equations, results and sections without the prefix S are to the main paper.

\section{Additional results on correction, cost and fragmentation}\label{S-app:additional}

\textbf{Robustness of the correction rule} (Section~\ref{sec:correction}).

\begin{proposition}[Robustness of the coherence gap]\label{S-prop:robust}
\econ (a) If each intermediary's response is nonlinear but responsive (stronger discrepancies call forth stronger effort, within bounds), amplification is finite if and only if the gap is positive, and it is bounded above by an expression inversely proportional to the gap. In growing economies whose checks each involve a bounded number of links, a gap shrinking to zero still produces unbounded amplification. (b) If intermediaries coordinate, the exact result of Theorem~\ref{thm:1} survives up to the strongest and weakest weights.
\formal (a) Replace $y=\kappa Dz$ by $y_\tau=\kappa\varphi((Dz)_\tau)$, where $\varphi(0)=0$ is strongly monotone with modulus $\mu>0$ and Lipschitz with constant $M$. Then
\[
\frac{1}{\kappa M\sqrt{\Gamma(X)\,\lambda_{\max}(H)}}\ \le\ A(X)\ \le\ \frac{1}{\kappa\mu\,\Gamma(X)},
\]
and $A(X)=+\infty$ if $\Gamma(X)=0$. If $\Gamma(Y_n)\ge\gamma>0$ then $\sup_nA(Y_n)\le1/(\kappa\mu\gamma)$. If each link lies in at most $\rho$ checks, then $\lambda_{\max}(H)\le3\rho$, so for triangular patches ($\rho=2$) $A\ge\sqrt{L/18}/(\kappa M)$, which is unbounded. (b) If $y=\kappa KDz$ for a symmetric positive definite $K$ with eigenvalues in $[k_{\min},k_{\max}]$, then
\[
\frac{1}{\kappa k_{\max}\Gamma(X)}\ \le\ A(X)\ \le\ \frac{1}{\kappa k_{\min}\Gamma(X)} .
\]
\end{proposition}

\textbf{Monitoring expenditure and regional fragmentation} (Section~\ref{sec:arch}).

\subsection*{Proofs}

\medskip
\noindent\emph{Proof of Proposition~\ref{S-prop:robust}.} (b) The dynamics are $\dot z=-\kappa D^*KDz+\xi$ with $D^*KD$ symmetric positive semidefinite, so the proof of Theorem~\ref{thm:1} gives $A=1/(\kappa\lambda_{\min}(D^*KD))$. Since $k_{\min}H\preceq D^*KD\preceq k_{\max}H$, $\lambda_{\min}(D^*KD)\in[k_{\min}\Gamma,k_{\max}\Gamma]$. (a) Upper bound: with $V=\tfrac12\|z\|^2$ and $\|\xi\|\le1$, $\dot V=-\kappa\langle Dz,\varphi(Dz)\rangle+\langle z,\xi\rangle\le-\kappa\mu\|Dz\|^2+\|z\|\le-\kappa\mu\Gamma\|z\|^2+\|z\|$, which is negative whenever $\|z\|>1/(\kappa\mu\Gamma)$. Lower bound: let $w\in C$ be a unit eigenvector of $H$ for $\Gamma$, so $\|Dw\|=\sqrt\Gamma$, and take $\xi\equiv w$. Strong monotonicity gives a unique $z^\ast$ solving $\kappa D^*\varphi(Dz^\ast)=w$, and $\tfrac{d}{dt}\tfrac12\|z-z^\ast\|^2\le-\kappa\mu\Gamma\|z-z^\ast\|^2$, so $z(t)\to z^\ast$ and $A\ge\|z^\ast\|$. Taking the inner product of the equation with $w$ gives $1=\kappa\langle Dw,\varphi(Dz^\ast)\rangle\le\kappa\sqrt\Gamma\,M\|Dz^\ast\|\le\kappa\sqrt\Gamma\,M\sqrt{\lambda_{\max}}\|z^\ast\|$. If $\Gamma=0$ then $Dw=0$ and $\tfrac{d}{dt}\langle z,w\rangle=1$, so $\|z(t)\|\ge t$. Finally $\|d_1\|^2\le\|d_1\|_1\|d_1\|_\infty\le3\rho$ because each row has three entries of modulus one and each column at most $\rho$, and $\Gamma\le3/L$ with $\rho=2$ gives $A\ge1/(\kappa M\sqrt{3\cdot6/L})$. A numerical check on triangular patches of sides 4, 6 and 8 confirms both bounds. \hfill$\square$

\noindent\emph{Heterogeneous capacities.} If intermediary $\tau$ has capacity $\kappa_\tau\in[\kappa_{\min},\kappa_{\max}]$, the dynamics become $\dot z=-D^*KDz+\xi$ with $K=\operatorname{diag}(\kappa_\tau)$. Since $\kappa_{\min}H\preceq D^*KD\preceq\kappa_{\max}H$, the smallest eigenvalue of $D^*KD$ lies in $[\kappa_{\min}\Gamma,\kappa_{\max}\Gamma]$, and the argument of Theorem~\ref{thm:1} gives $1/(\kappa_{\max}\Gamma)\le A(X)\le1/(\kappa_{\min}\Gamma)$.
\medskip

\medskip
\begin{proposition}[The cost of regional fragmentation]\label{S-prop:3}
\econ If checks operate only inside regional blocs, every bilateral link between blocs adds an inconsistency direction that no check monitors, except for a small correction for the number of blocs. Each additional cross-bloc check closes at most one such direction.
\formal
Partition the connected bilateral graph into $k$ connected induced regions with $M$ links between regions. If all checks are confined to individual regions, then $\dim\H^1(X;\R)\ge M-k+1$. Adding $s$ cross-regional checks reduces the lower bound by at most $s$. If each region has between $\nu$ and $B\le n/2$ vertices, and $|\partial S|\ge h|S|$ for all $|S|\le n/2$, then $\dim\H^1(X;\R)\ge(h/2-1/\nu)\,n+1-s$.
\end{proposition}

\noindent\emph{Proof.} Set all within-region link values to zero and allow arbitrary values on the $M$ inter-regional links. No regional check detects this $M$-dimensional space. The gradients lying in it are exactly those from potentials constant within each connected region, of dimension $k-1$. Modding them out leaves $M-k+1$ undetected directions. Each additional independent cross-regional check removes at most one dimension. Summing the graph expansion inequalities across regions gives $2M\ge hn$, and $k\le n/\nu$.

\subsection*{The monitoring burden and the cost of a target}

\begin{proposition}[The monitoring burden of bilateral integration]\label{S-prop:1}
\econ In a connected network of $m$ bilateral agreements among $n$ participants, at least $m-n+1$ independent multilateral checks are needed before every inconsistency can be detected. Each new bilateral agreement that no existing check covers adds one undetected direction of inconsistency.
\formal $\dim \H^1(X;\R)\ge m-n+1-|T|$, and a new link that enters no existing check raises $\dim\H^1$ by exactly one.
\end{proposition}

A designer can meet a target $A(X)\le\bar A$, for a given $\bar A>0$, by buying capacity or by arranging checks better. Let each link cost $c_E$ and each check $c_T+c_\kappa\kappa$.

\begin{proposition}[Cost of a target]\label{S-prop:cost}
\econ Architecture substitutes for capacity. In local meshes the capacity each check needs grows with the square root of the economy's size, so total cost rises faster than the economy; in the sparse architecture it is constant and total cost is proportional to the economy.
\formal (i) The cheapest capacity is $\kappa=1/(\bar A\,\Gamma(X))$, with total cost $c_Em+\big(c_T+c_\kappa/(\bar A\,\Gamma(X))\big)|T|$. (ii) For a triangular patch of side $L$, total cost is at least $c_\kappa L^3/(3\bar A)$, of order $n^{3/2}$. (iii) For $Y_n$, total cost is $\big[73c_E+219\big(c_T+c_\kappa/(\bar A(9-4\sqrt2))\big)\big]|V(Y_n)|$.
\end{proposition}

\subsection*{Proofs of the results in Sections 2 to 4}

Notation is as in Appendix~\ref{app:proofs} of the main paper.

\medskip
\noindent\emph{Proof of Proposition~\ref{S-prop:1}.} Since the bilateral graph is connected, $\operatorname{rank}d_0=n-1$ and $\dim C=m-n+1$. Because $\operatorname{rank}d_1\le|T|$ and $d_1d_0=0$, $\dim\H^1(X;\R)=m-\operatorname{rank}d_1-(n-1)\ge m-n+1-|T|$. A new unchecked link contributes a zero column to $d_1$, leaves its rank unchanged, and enlarges the cycle dimension by one.

\medskip
\noindent\emph{Proof of Theorem~\ref{thm:1}.} For positive definite $H$ the solution of \eqref{eq:5} is
\[z(t)=\int_0^t e^{-\kappa H(t-s)}\xi(s)\,ds.\] The spectral theorem yields $\|z(t)\|\le\int_0^t e^{-\kappa\Gamma(t-s)}ds\le 1/(\kappa\Gamma)$. A constant disturbance equal to a normalized smallest-eigenvalue eigenvector attains the limit. If $H$ has a null vector $w$ with $\|w\|=1$, choosing $\xi(t)=w$ yields $z(t)=tw$, so amplification is infinite.

\medskip
\noindent\emph{Proof of Theorem~\ref{thm:2}.} On a triangulated disk, $\H^1(X;\R)=\H^2(X;\R)=0$. Euler's formula implies $\dim C=t$, so the face-boundary operator $\partial_2:\R^T\to C$ is invertible and $D=\partial_2^*|_C$. The smallest eigenvalues of $D^*D$ and $\partial_2^*\partial_2$ are therefore equal and strictly positive. Give the faces consistent orientation, and let $f\in\R^T$ assign one to every face. Interior links cancel in $\partial_2f$; the boundary links each have unit coefficient. Hence $\|\partial_2f\|^2=b_\partial$ while $\|f\|^2=t$, and the Rayleigh principle gives $\Gamma\le b_\partial/t$.

\medskip
\noindent\emph{Proof of Theorem~\ref{thm:3}.} Fix the residue field size $q_0=8$ and consider the two-dimensional Bruhat--Tits building $\mathcal B$ associated with $\mathrm{PGL}_3(\F_{q_0}((t)))$. The explicit constructions of \citet{lsv2005a,lsv2005b} provide arbitrarily large finite quotients of this building. We take sufficiently deep quotients so that the building's vertices, links, triangles and vertex links descend without identifications inside a closed star. These are connected finite simplicial complexes, all of whose vertex links are incidence graphs of the projective plane $\mathbb P^2(\F_{q_0})$.

The incidence graph has $2(q_0^2+q_0+1)$ vertices, degree $q_0+1$, and adjacency eigenvalues $\pm(q_0+1)$ and $\pm\sqrt{q_0}$. Its smallest nonzero normalized-Laplacian eigenvalue is therefore $\varepsilon_{q_0}=1-\sqrt{q_0}/(q_0+1)$. Garland's local-to-global inequality for a two-dimensional complex gives the lower bound $2\varepsilon_{q_0}-1$ for the weighted real upper-cochain spectral gap on 1-cochains orthogonal to exact gradients \citep[Theorem~2.5 and the discussion of Garland's method therein]{lubotzky2018}. In these quotients every link belongs to exactly $q_0+1$ triangular faces and every face has weight one. Thus the weighted link norm is $(q_0+1)$ times the unweighted Euclidean norm while the face norm is unchanged, and $\operatorname{im}d_0$ has the same orthogonal complement in either inner product. Converting to the unweighted convention in \eqref{eq:3} yields
\[
\Gamma(Y_n)\ \ge\ (q_0+1)(2\varepsilon_{q_0}-1)=q_0+1-2\sqrt{q_0}=(\sqrt{q_0}-1)^2 .
\]
For $q_0=8$ this is $9-4\sqrt2>0$. The positive bound also establishes $\H^1(Y_n;\R)=0$, so no zero modes were discarded in passing to the upper spectrum.

Every vertex has $2(q_0^2+q_0+1)=146$ neighbors. The handshake lemma gives $|E|=(q_0^2+q_0+1)|V|=73|V|$. Each link belongs to $q_0+1=9$ triangles and each triangle has three links, so $|T|=(q_0+1)|E|/3=219|V|$. Both the vertex degree and the link--face incidence are therefore bounded uniformly in the size of the quotient. With unit bilateral-link cost $c_E>0$ and triangular-check cost $c_T>0$, total cost is exactly $(73c_E+219c_T)|V|$. Theorem~\ref{thm:1} gives the adjustment bound; for disk families, Theorems~\ref{thm:2} and~\ref{thm:1} give divergence. The quotients show that the feasibility conditions can be met; their high local degrees and triangle memberships are not a proposal for an actual treaty system.

\medskip
\noindent\emph{Proof of Proposition~\ref{S-prop:cost}.} (i) By Theorem~\ref{thm:1}, $A(X)=1/(\kappa\Gamma(X))\le\bar A$ if and only if $\kappa\ge 1/(\bar A\Gamma(X))$; capacity cost is increasing in $\kappa$, so the minimum is attained at equality, and each of the $|T|$ checks pays $c_T+c_\kappa\kappa$. (ii) By Theorem~\ref{thm:2}, $\Gamma\le b_\partial/t=3/L$, so $\kappa\ge L/(3\bar A)$. With $t=L^2$ checks, capacity cost alone is at least $L^2\cdot c_\kappa L/(3\bar A)$. Since $(L+1)(L+2)\le 6L^2$ for $L\ge1$, $n\le 3L^2$ and $L\ge\sqrt{n/3}$, so cost is $\Omega(n^{3/2})$. (iii) Substitute $|E|=73|V|$, $|T|=219|V|$ and $\Gamma\ge9-4\sqrt2$ into (i).
\medskip

\medskip
\noindent\emph{Proof of Proposition~\ref{prop:micro}.} (a) An arbitrageur on $\tau$ maximizes $y_\tau(Dz)_\tau-y_\tau^2/(2\kappa)$, so $y_\tau=\kappa(Dz)_\tau$ and the profit is $\kappa(Dz)_\tau^2/2$; summing gives $\frac\kappa2\|Dz\|^2$. A unit round trip around $\tau$ moves the link prices by $-b_\tau$, so the flows move $z$ by $-\sum_\tau y_\tau b_\tau=-D^*y=-\kappa Hz$ per unit of time; adding the shocks gives the stated equation. (b) For $\Gamma>0$ the Ornstein--Uhlenbeck process $dz=-\kappa Hz\,dt+\sigma\,dW$ on $C$ has a stationary covariance solving the Lyapunov equation $\kappa H\Sigma+\kappa\Sigma H=\sigma^2I_C$, whose unique solution is $\Sigma=\frac{\sigma^2}{2\kappa}H^{-1}$. Country $i$ bears half the loss $\frac\eta2z_e^2$ on each of its links, so its expected loss is $\frac\eta4\sum_{e\ni i}\mathbb E z_e^2=\frac\eta4\sum_{e\ni i}\Sigma_{ee}$. (c) Expected rents are $\frac\kappa2\mathbb E\|Dz\|^2=\frac\kappa2\operatorname{tr}(D\Sigma D^*)=\frac\kappa2\operatorname{tr}(H\Sigma)=\frac\kappa2\cdot\frac{\sigma^2}{2\kappa}\dim C=\frac{\sigma^2}{4}r$. (d) After adding $\tau$, the rents on $\tau$ are $\frac\kappa2\mathbb E(b^\top z)^2=\frac\kappa2\cdot\frac{\sigma^2}{2\kappa}\,b^\top(H+bb^\top)^{-1}b=\frac{\sigma^2}{4}\frac{q}{1+q}$ by the Sherman--Morrison formula. Total rents are unchanged by (c), so the rents on the other checks fall by the same amount. \hfill$\square$

\noindent\emph{Proof of Proposition~\ref{prop:local}.} Give every check weight one and every link weight $t(e)$, the number of checks containing $e$. Let $B=d_1^*d_1$ and $M=\operatorname{diag}(t(e))$. The weighted partner graph of $v$ is $\mathrm{Lk}(v)$ with the weights of the normalized Laplacian in the hypothesis. Garland's local-to-global inequality for weighted two-dimensional complexes \citep{lubotzky2018}, the inequality used for Theorem~\ref{thm:3}, says that if every $\mathrm{Lk}(v)$ is connected with second eigenvalue at least $\lambda>1/2$, then $M^{-1}B$ has smallest eigenvalue at least $2\lambda-1$ on the $M$-orthogonal complement of $\operatorname{im}d_0$. Hence $\ker d_1=\operatorname{im}d_0$, so $\dim\H^1(X;\R)=0$, and the eigenvalue zero has multiplicity $n-1$ both for $B$ and for $M^{-1}B$. The $n$th smallest eigenvalue of $B$ is then $\Gamma(X)$. The $n$th smallest eigenvalue of $M^{-1}B$ is at least $2\lambda-1$. Since $M\succeq t_{\min}I$, we have $z^*Bz/z^*z\ge t_{\min}\,z^*Bz/z^*Mz$ for every $z$. The Courant--Fischer min--max formula then gives $\Gamma(X)\ge t_{\min}(2\lambda-1)$. Because every link is in a check, $t_{\min}\ge1$. The amplification bound is Theorem~\ref{thm:1}. If every participant has at most $\bar d$ partners, then $|E|\le\bar dn/2$. Each check is a pair of partners of each of its three nodes, so $|T|\le\frac13 n\binom{\bar d}2\le\bar d^2n/6$. The cost formula of Proposition~\ref{S-prop:cost}(i) with $\Gamma\ge2\lambda-1$ gives the cost bound.

\section{The Shapley value of the coherence game}\label{S-app:shapley}

The core of the coherence game (Corollary~\ref{thm:core}) gives everything to the indispensable countries. A value spreads the gains according to each country's contribution. Let $v(K)=\Gamma(T\cup A(K))-\Gamma(T)$ be the gain in coherence that coalition $K$ can achieve, $\phi_x$ the Shapley value of country $x$ in the game $v$ \citep{shapley1953}, and $\mathrm{SS}_x(\gamma)$ its Shapley--Shubik index in the coherence game with target $\gamma$ \citep{shapleyshubik1954}, the probability that $x$ turns a losing coalition into a winning one when countries arrive in random order.

\begin{proposition}[The Shapley value of the coherence game]\label{S-prop:shapley}
\econ A country's Shapley value of the gains from coherence is its power to complete a winning coalition, averaged over all targets. A country that Theorem~\ref{thm:veto} certifies as indispensable for a range of targets receives at least a fixed fraction of that range, so the value, like the core, rewards indispensable countries, but it also rewards countries that are pivotal without being indispensable.
\formal (a) $\phi_x=\int_{\Gamma(T)}^{\bar\gamma}\mathrm{SS}_x(\gamma)\,d\gamma$. (b) Under the conditions of Theorem~\ref{thm:veto} (with $\hat\lambda_x=\infty$ otherwise), $\phi_x\ge\big(\bar\gamma-\max\{\hat\lambda_x,\Gamma(T)\}\big)^+/n$. (c) With $c_x=\Gamma(T\cup A(V\setminus\{x\}))$, $\mathrm{Vet}_\gamma=\{y:c_y<\gamma\}$ and $\omega_\gamma$ the size of the smallest coalition that reaches $\gamma$,
\[
\int_{c_x}^{\bar\gamma}\frac{d\gamma}{\omega_\gamma}\ \le\ \phi_x\ \le\ \int_{c_x}^{\bar\gamma}\frac{d\gamma}{|\mathrm{Vet}_\gamma|}+\int_{\Gamma(T)}^{c_x}\frac{d\gamma}{|\mathrm{Vet}_\gamma|+1}.
\]
\end{proposition}

By \citet{sterntettenhorst2019}, $\phi_x$ is also the value of the grand coalition in $x$'s component of the Hodge decomposition of $v$ on the cube of coalitions. The game itself is defined by the Hodge theory of the checks, so the value of a country is the outcome of two Hodge decompositions, one of the agreements and one of the coalitions.

\medskip
\noindent\emph{Proof of Proposition~\ref{S-prop:shapley}.} (a) Adding checks never lowers the gap, so $\Gamma(T)\le\Gamma(T\cup A(K))\le\bar\gamma$ and $v(K)=\int_{\Gamma(T)}^{\bar\gamma}\mathbf 1[\Gamma(T\cup A(K))\ge\gamma]\,d\gamma$. The integrand is the characteristic function of the coherence game with target $\gamma$, and the Shapley value is linear in the game and equals the Shapley--Shubik index on simple games. (b) For $\gamma>\max\{\hat\lambda_x,\Gamma(T)\}$, Theorem~\ref{thm:veto} with $A=A(V\setminus\{x\})$ gives $\Gamma(T\cup A(V\setminus\{x\}))\le\hat\lambda_x<\gamma$, so $x\in\mathrm{Vet}_\gamma$. In the orders in which $x$ arrives last, which have probability $1/n$, the coalition before $x$ loses and the grand coalition wins, so $\mathrm{SS}_x(\gamma)\ge1/n$. Integrate over $\gamma$ and use (a). (c) For $\gamma>c_x$, $x$ is a veto player, so every winning coalition contains $x$; if $x$ arrives after the other members of a smallest winning coalition, which has probability $1/\omega_\gamma$, the coalition before $x$ loses and the one including $x$ wins, so $\mathrm{SS}_x(\gamma)\ge1/\omega_\gamma$. Conversely, $x$ can be pivotal only if the coalition including it contains every veto player, and only if the coalition before it does when $x$ is not one; $x$ arrives after all other veto players with probability $1/|\mathrm{Vet}_\gamma|$ if $x\in\mathrm{Vet}_\gamma$ and $1/(|\mathrm{Vet}_\gamma|+1)$ otherwise. Integrate over $\gamma$ and use (a). \hfill$\square$

In the treaty network (Section~\ref{sec:bargevidence}), estimated from 120 random orders, the ten economies of Table~\ref{tab:veto} receive 51 percent of the Shapley value, Ukraine alone 9 percent and Germany almost nothing; across all 80 economies the Shapley value has a rank correlation of $-0.88$ with the gap reached without the economy and of $0.15$ with its number of partners. Kazakhstan's Shapley share is 3.7 percent, against 0.3 percent of the package's value. Part (c), with the smallest winning coalitions bounded by the greedy coalitions of Section~\ref{sec:bargevidence}, gives lower bounds between a third and 70 percent of the estimated values for the economies of Table~\ref{tab:veto}; the bounds of part (c) have a rank correlation of 0.88 with the estimates. The upper bound is loose for small targets, at which no economy is indispensable.

\section{Two benchmarks}\label{S-app:bench}

This appendix states the two benchmark economies summarized in Section~\ref{sec:grid}; their proofs follow in Appendices~\ref{S-app:grid} and~\ref{S-app:symm}.

\textbf{Geography: agreements on a grid.} Place $n=L^2$ participants on a grid, wrapped into a torus, each with agreements with its eight nearest neighbours. Each square cell then contains four triangles whose links exist, any three of which check every inconsistency in the cell. Let $T_0$ check the same three triangles in every cell, and let the fourth be the candidate.\footnote{The torus has two global loops that no triangle detects; shocks are restricted to $W=\operatorname{im}d_1^*$ and $H^{-1}$ is replaced by its pseudo-inverse on $W$, which contains every candidate. On a flat grid the loops do not exist and the numbers behave the same way.}

\begin{proposition}[Grid economies]\label{S-prop:grid}
\econ In a grid of local agreements, a missing check is worth more the larger the economy, while its members' gain stays bounded. For any cost above that bound, the initial architecture is stable but not efficient once the economy is large enough.
\formal Every candidate satisfies $S_L(\tau)=\frac{3}{\pi}\log L+O(1)$, $s_L(\tau)S_L(\tau)\le\bar P$ and $s_L(\tau)=O(1/\log L)$, for a constant $\bar P$, and $\frac{16}{81L^2}\le\Gamma\le\frac{2\pi^2}{L^2}$. For every $c>\bar P$ there is $\bar L$ such that for $L\ge\bar L$, $T_0$ is stable and adding any candidate lowers the planner's objective by $S_L(\tau)-c>0$.
\end{proposition}

A Fourier transform reduces the problem to a $3\times3$ matrix at each frequency $k\in[-\pi,\pi]^2$, whose determinant is the Laplacian of the grid's neighbour graph and vanishes like $3|k|^2$ near zero. The check's response therefore decays slowly with distance: its value is a sum of $12/|k|^2$ over frequencies, which grows like $\log L$, while its effect on its members' links converges (Appendix~\ref{S-app:grid}).\footnote{Numerically, $S_L-\frac3\pi\log L$ converges to $1.075$ and the members' gain to $2.048$; at $L=1{,}024$ the share is $0.27$ and still falling. On three-dimensional lattices the sum stays bounded, and in long thin networks weak directions do not arise.}

\textbf{No geography: symmetric economies.} Let $n$ be odd, label participants by $\Z_n$, let every pair have an agreement, and let the architecture be invariant under rotation: if $\{x,y,z\}$ is checked, so is $\{x+1,y+1,z+1\}$. A link $\{p,p+d\}$, $1\le d\le(n-1)/2$, has \emph{type} $d$, so an inconsistency is a family of functions $z_d:\Z_n\to\R$. Rotation commutes with $H$, so the Fourier transform $\hat z_d(k)=\sum_pz_d(p)e^{2\pi ikp/n}$, $k=0,\dots,n-1$, splits $H$ into $n$ Hermitian blocks $\hat H(k)$ of size $(n-1)/2$. Let $K_\mu$ be the set of frequencies $k$ at which $\hat H(k)$ has an eigenvalue below $\mu$ on inconsistencies, that is, ignoring the one null direction at each $k\neq0$ that comes from benchmarks.

\begin{proposition}[Symmetric economies]\label{S-prop:symm}
\econ When no country is special, the weak directions are waves running through the whole economy. A wave made of few frequencies cannot be concentrated on three countries, so the members of a check that repairs it receive about $3/n$ of its value per weak frequency. The wedge widens in proportion to the size of the economy.
\formal Let the architecture be rotation-invariant with $\Gamma>0$, and $\mu>0$. For every candidate,
\begin{equation}\label{S-eq:symmbound}
s(\tau)\le3\Big(\sqrt{\tfrac{|K_\mu|}{n}}+\frac{\sqrt3}{\mu\sqrt{S(\tau)}}\Big)^2,\qquad
s(\tau)S(\tau)\le3\Big(\sqrt{\tfrac{|K_\mu|\,S(\tau)}{n}}+\frac{\sqrt3}{\mu}\Big)^2 .
\end{equation}
If $\mu\ge6/\sqrt c$, every candidate with $c\le S(\tau)<c\,n/(12|K_\mu|)$ is worth adding but is not created.
\end{proposition}

The proof (Appendix~\ref{S-app:symm}) is an uncertainty principle on $\Z_n$: a pattern whose Fourier transform is supported on a set $K$ of frequencies places at most a fraction $2|K|/n$ of its energy on any one country's links. The bound is sharp. A weak direction at $k=0$ is a rotation-invariant circulation around the whole ring, the purest form of the loop behind Theorem~\ref{thm:2}. It puts the same weight $\ell_i=1/n$ on every country, so part (b) of Theorem~\ref{thm:general} gives a share close to $3/n$. Weakening one orbit of checks so that $\Gamma\to0$, the members' share of the most valuable candidate converges to $3/n$ in every case we computed, while its value grows like $1/\Gamma$. Conversely, when weak directions occupy many frequencies they can combine into patterns concentrated on a few countries, and members keep most of the value.\footnote{When a single orbit of checks is weak at every frequency, members keep about 98 percent. Both benchmarks are sharp cases of Theorem~\ref{thm:general}: under rotation or translation invariance every country holds the same mass $\rho_\mu(i)=2N_\mu/n$ of the weak band, so there are no hubs.} Weaknesses that are local can be repaired privately; weaknesses that are global cannot.

\section{Proof of Proposition~\ref{S-prop:grid}}\label{S-app:grid}

\textbf{Setup.} Participants are $(x,y)\in\Z_L^2$. The cell with lower-left corner $(x,y)$ owns four links, oriented as written:
\[
h=(x,y)(x{+}1,y),\ \ v=(x,y)(x,y{+}1),\ \ d=(x,y)(x{+}1,y{+}1),\ \ a=(x{+}1,y)(x,y{+}1).
\]
Write $p_{00},p_{10},p_{01},p_{11}$ for the corners. The four triangles of the cell have boundaries
\begin{align*}
T_1&=\{p_{00},p_{10},p_{11}\}: & b_1&=h+v(x{+}1,y)-d,\\
T_2&=\{p_{00},p_{01},p_{11}\}: & b_2&=v+h(x,y{+}1)-d,\\
T_3&=\{p_{00},p_{10},p_{01}\}: & b_3&=h+a-v,\\
T_4&=\{p_{10},p_{01},p_{11}\}: & b_4&=a+h(x,y{+}1)-v(x{+}1,y),
\end{align*}
and $b_4=b_2+b_3-b_1$. $T_0$ contains $T_1,T_2,T_3$ in every cell; the candidate is $T_4$ in cell $(0,0)$, with members $(1,0)$, $(0,1)$ and $(1,1)$.

\textbf{Fourier transform.} For $k\in K_L=\frac{2\pi}{L}\Z_L^2$ and a link function $z$, let $\hat z(k)\in\C^4$ have entries $\hat z_t(k)=\sum_{(x,y)}z_t(x,y)e^{i(k_1x+k_2y)}$, $t\in\{h,v,d,a\}$, where $z_t(x,y)$ is the value on the type-$t$ link of cell $(x,y)$. Then $z_t(x,y)=L^{-2}\sum_k\hat z_t(k)e^{-i(k_1x+k_2y)}$ and $\langle z,w\rangle=L^{-2}\sum_k\hat z(k)^*\hat w(k)$. With $\xi=e^{ik_1}$ and $\eta=e^{ik_2}$, the boundary of the type-$j$ triangle of cell $(x_0,y_0)$ transforms to $e^{ik\cdot(x_0,y_0)}\beta_j(k)$, where
\begin{gather*}
\beta_1=(1,\xi,-1,0),\ \ \beta_2=(\eta,1,-1,0),\ \ \beta_3=(1,-1,0,1),\\ \beta_4=(\eta,-\xi,0,1)=\beta_2+\beta_3-\beta_1 .
\end{gather*}
Let $B(k)=[\beta_1\ \beta_2\ \beta_3]$, $G(k)=B(k)^*B(k)$ and $\alpha=(-1,1,1)^\top$, so $\beta_4=B\alpha$. Since $H=\sum_{\tau\in T_0}b_\tau b_\tau^\top$ and $\sum_{(x_0,y_0)}e^{i(k-k')\cdot(x_0,y_0)}=L^2\mathbf 1\{k=k'\}$,
\[
\widehat{Hz}(k)=\hat H(k)\hat z(k),\qquad \hat H(k)=B(k)B(k)^* .
\]

\textbf{The $3\times3$ matrix.} Write $c_1=\cos k_1$, $c_2=\cos k_2$, $c_\pm=\cos(k_1\pm k_2)$. The entries of $G$ are $G_{jj}=3$, $g_{12}=1+\bar\xi+\eta$, $g_{13}=1-\bar\xi$ and $g_{23}=\bar\eta-1$, with $|g_{12}|^2=3+2c_1+2c_2+2c_+$, $|g_{13}|^2=2-2c_1$, $|g_{23}|^2=2-2c_2$ and $\operatorname{Re}(g_{12}g_{23}\bar g_{13})=1-c_1-c_2+2c_+-c_-$. Hence
\[
\det G=27+2\operatorname{Re}(g_{12}g_{23}\bar g_{13})-3\big(|g_{12}|^2+|g_{13}|^2+|g_{23}|^2\big)=8-2c_1-2c_2-2c_+-2c_-,
\]
which is the symbol of the Laplacian of the grid's neighbour graph and equals $(2-2c_1)+(2-2c_2)+(2-2c_+)+(2-2c_-)$. Using $2-2\cos t\ge4t^2/\pi^2$ for $|t|\le\pi$ on the first two terms and $2-2\cos t=t^2+O(t^4)$,
\begin{equation}\label{S-eq:detG}
\det G(k)\ge\frac{4}{\pi^2}|k|^2\ \ \text{on }[-\pi,\pi]^2,\qquad \det G(k)=3|k|^2+O(|k|^4).
\end{equation}
The adjugate gives $\alpha^*\operatorname{adj}(G)\alpha=\sum_j\operatorname{adj}(G)_{jj}+2\operatorname{Re}\big(3g_{12}-g_{13}\bar g_{23}-g_{12}g_{23}+3g_{13}-3g_{23}+g_{13}\bar g_{12}\big)$. The principal minors sum to $20-2c_+$ and the real part equals $10+c_--2c_+$, so
\begin{equation}\label{S-eq:adjG}
N(k):=\alpha^*\operatorname{adj}(G(k))\,\alpha=40-6c_++2c_-\in[32,48],\qquad N(0)=36 .
\end{equation}

\textbf{The response of the candidate.} For $k\neq0$, $\det G(k)>0$, so $B(k)$ has full column rank, $\hat H(k)^+B=B(B^*B)^{-1}$ and $\hat u(k)=\hat H(k)^+\beta_4=B(k)G(k)^{-1}\alpha$. Then $\beta_4^*\hat H^+\beta_4=\alpha^*\alpha=3$ and
\[
\|\hat u(k)\|^2=\alpha^*G^{-1}B^*BG^{-1}\alpha=\alpha^*G(k)^{-1}\alpha=f(k):=\frac{N(k)}{\det G(k)} .
\]
At $k=0$, $\beta_1(0)=\beta_2(0)=(1,1,-1,0)$ is orthogonal to $\beta_3(0)=\beta_4(0)=(1,-1,0,1)$, so $\hat H(0)$ has rank two and $\hat u(0)=\beta_3(0)/3$. The kernel of $\hat H(0)$ consists of the two global loops; $\hat H(k)$ has a one-dimensional kernel, the gradients, for $k\neq0$. By Parseval,
\[
q=\frac{3(L^2-1)+1}{L^2}=3-\frac{2}{L^2},\qquad \|u\|^2=\frac{1}{L^2}\Big(\frac13+\sum_{k\neq0}f(k)\Big).
\]

\textbf{Two lattice sums.} For $R\ge1$,
\begin{equation}\label{S-eq:lattice}
\sum_{0<\|m\|_\infty\le R}\frac{1}{|m|^2}=2\pi\log R+O(1),\qquad \sum_{0<\|m\|_\infty\le R}\frac{1}{|m|}=O(R),
\end{equation}
with $m\in\Z^2$. For the first, let $Q_m$ be the unit square centred at $m$. For $|m|\ge2$, the gradient of $|x|^{-2}$ is $O(|m|^{-3})$ on $Q_m$, so $\big||m|^{-2}-\int_{Q_m}|x|^{-2}dx\big|=O(|m|^{-3})$, and $\sum|m|^{-3}<\infty$. The union of the $Q_m$ is the square $[-R-\frac12,R+\frac12]^2$ minus $Q_0$, on which $\int|x|^{-2}dx=2\pi\log R+O(1)$, because the region contains the annulus $\{1\le|x|\le R\}$ and is contained in $\{\frac12\le|x|\le\sqrt2(R+1)\}$. The second sum is bounded by the same comparison with $\int|x|^{-1}dx$.

\textbf{Social value.} By \eqref{S-eq:detG} and \eqref{S-eq:adjG}, $f(k)=12/|k|^2+g(k)$ with $g$ bounded on $[-\pi,\pi]^2\setminus\{0\}$: near $0$, $f=(36+O(|k|^2))/(3|k|^2+O(|k|^4))$, and away from $0$ both $f$ and $12/|k|^2$ are bounded. Writing $k=2\pi m/L$ and using \eqref{S-eq:lattice} with $R=L/2$,
\[
\frac{1}{L^2}\sum_{k\neq0}f(k)=\frac{3}{\pi^2}\sum_{0<\|m\|_\infty\le L/2}\frac{1}{|m|^2}+O(1)=\frac{6}{\pi}\log L+O(1).
\]
Hence $\|u\|^2=\frac6\pi\log L+O(1)$ and $S_L=2\|u\|^2/(1+q)=\|u\|^2/(2-L^{-2})=\frac3\pi\log L+O(1)$.

\textbf{Members' gain.} By \eqref{S-eq:detG} and \eqref{S-eq:adjG}, $|\hat u_t(k)|\le\sqrt{f(k)}\le\pi\sqrt{12}/|k|$ for $k\neq0$, and $|\hat u_t(0)|\le1/\sqrt3$. For every link $e$,
\[
|u_e|\le\frac{1}{L^2}\Big(\frac{1}{\sqrt3}+\sum_{k\neq0}\frac{\pi\sqrt{12}}{|k|}\Big)=\frac{1}{\sqrt3L^2}+\frac{\sqrt{12}}{2L}\sum_{0<\|m\|_\infty\le L/2}\frac{1}{|m|}\le U,
\]
with $U$ independent of $L$ by \eqref{S-eq:lattice}. The members' joint gain is $\sum_e\nu_e(\tau)u_e^2/(1+q)$, the three members have eight links each, so $\sum_e\nu_e(\tau)=24$, and $1+q\ge3$. Hence $s_LS_L\le\bar P:=8U^2$, and $s_L=O(1/\log L)$.

\textbf{Coherence gap.} On $W$, the spectrum of $H$ is the union over $k$ of the nonzero eigenvalues of $\hat H(k)=BB^*$, which coincide with the eigenvalues of $G(k)=B^*B$ for $k\neq0$ and equal $\{3,6\}$ at $k=0$. For $n=(1,-1,0)$, $n^*G(k)n=(2-2c_1)+(2-2c_2)\le|k|^2$, so $\lambda_{\min}(G(k))\le|k|^2/2$, and $k=(2\pi/L,0)$ gives $\Gamma\le2\pi^2/L^2$. Conversely, $\lambda_{\min}(G)\ge\det G/\lambda_{\max}(G)^2\ge\det G/(\operatorname{tr}G)^2=\det G/81\ge\frac{4}{81\pi^2}|k|^2\ge\frac{16}{81L^2}$ for $k\neq0$.

\textbf{Stability without efficiency.} If $c>\bar P$, no group of three gains from adding any candidate, and $T_0$ contains no added checks to drop, so $T_0$ is stable. Once $L$ is large enough that $S_L>c$, adding any one candidate lowers the planner's objective by $S_L-c$, by Proposition~\ref{prop:wedge}. \hfill$\square$

\smallskip
\noindent\emph{Numerical check.} The replication package compares these formulas with brute-force computation on the torus for $L=4,\dots,14$ (agreement to machine precision) and evaluates them for $L$ up to $1{,}024$, where $S_L-\frac3\pi\log L=1.075$ and the members' gain is $2.048$.

\section{Proof of Proposition~\ref{S-prop:symm}}\label{S-app:symm}

\textbf{Fourier transform.} Identify $\R^E$ with families $z=(z_d)_{d=1}^{\ell}$, $\ell=(n-1)/2$, of functions $z_d:\Z_n\to\R$, where $z_d(p)$ is the value on the link $\{p,p+d\}$ oriented from $p$ to $p+d$. Since $n$ is odd, these are all the links, each counted once. With $\hat z_d(k)=\sum_pz_d(p)e^{2\pi ikp/n}$, Parseval's identity reads $\|z\|^2=\frac1n\sum_{k,d}|\hat z_d(k)|^2$. Rotation $R:z_d(p)\mapsto z_d(p-1)$ maps cycles to cycles and, because the architecture is rotation-invariant, commutes with $H=D^*D$. Hence $H$ acts on each frequency $k$ separately through a Hermitian $\ell\times\ell$ matrix $\hat H(k)$; for $k\neq0$ it has one null direction, the transform of the gradients $d_0x$, which are not cycles and are excluded. Because $H$ is real, $\hat H(-k)=\overline{\hat H(k)}$, so $K_\mu=-K_\mu$.

\textbf{Uncertainty step.} Let $z$ be real with $\hat z_d(k)=0$ for $k\notin K$. Then $z_d(p)=\frac1n\sum_{k\in K}\hat z_d(k)e^{-2\pi ikp/n}$, and the Cauchy--Schwarz inequality gives
\[|z_d(p)|^2\le\frac{|K|}{n^2}\sum_{k\in K}|\hat z_d(k)|^2.\] Country $i$ is an endpoint of exactly two links of each type, $\{i,i+d\}$ and $\{i-d,i\}$. Summing over them and over $d$, and using Parseval,
\[
\sum_{e\ni i}z_e^2\ \le\ \frac{2|K|}{n^2}\sum_{k,d}|\hat z_d(k)|^2\ =\ \frac{2|K|}{n}\,\|z\|^2 .
\]

\textbf{Splitting the response.} Let $P$ be the spectral projector of $H$ onto eigenvalues below $\mu$. Because $H$ is block diagonal in frequency, $P$ is too, and $u_K=Pu$ is real with Fourier support in $K_\mu$. Write $u=u_K+u_R$. Then $u_R=(I-P)H^{-1}b=H^{-1}(I-P)b$ has $\|u_R\|\le\|b\|/\mu=\sqrt3/\mu$, and $\|u_K\|\le\|u\|$. For each member $i$, by Minkowski's inequality and $\sum_{e\ni i}x_e^2\le\|x\|^2$,
\[
\Big(\sum_{e\ni i}u_e^2\Big)^{1/2}\le\Big(\frac{2|K_\mu|}{n}\Big)^{1/2}\|u\|+\frac{\sqrt3}{\mu}.
\]
By Proposition~\ref{prop:wedge}, $s(\tau)=\sum_{i\in\tau}\sum_{e\ni i}u_e^2\,/\,(2\|u\|^2)$. Summing the square of the last display over the three members,
\[
s(\tau)\le\frac32\Big(\sqrt{\tfrac{2|K_\mu|}{n}}+\frac{\sqrt3}{\mu\|u\|}\Big)^2 .
\]
Since $\|u\|^2=S(\tau)(1+q)/2\ge S(\tau)/2$, this gives the first inequality in~\eqref{S-eq:symmbound}; multiplying by $S(\tau)$ gives the second. The limit statement is immediate.

\textbf{The window.} If $\mu\ge6/\sqrt c$ then $\sqrt3/\mu\le\frac12\sqrt{c/3}$, and if $S(\tau)<cn/(12|K_\mu|)$ then $\sqrt{|K_\mu|S(\tau)/n}<\frac12\sqrt{c/3}$. The second inequality in~\eqref{S-eq:symmbound} then gives $s(\tau)S(\tau)<c$, so the members do not create $\tau$, while $S(\tau)\ge c$ means that adding it lowers the planner's objective (Proposition~\ref{prop:wedge}). \hfill$\square$

\smallskip
\noindent\emph{Numerical check.} The replication package verifies the block decomposition against dense computation on the complete graph for $n=7,9,11$, checks~\eqref{S-eq:symmbound} on every candidate in 16 random rotation-invariant architectures with $n\le41$, and computes the weak-orbit limit in which $n\,s(\tau)\to3$.

\section{Data construction and methods}\label{S-app:data}

\textbf{Sources.} The 80 economies are the OECD ICIO 2025 country list without the Rest of the World aggregate. DESTA 2.3 base treaties and coded accessions and withdrawals give year-specific memberships. Non-country and aggregate units are excluded or merged to national units.

\textbf{Cohomology.} For the connected treaty graph, $\dim\H^1(X;K)=|E|-|V|+1-\operatorname{rank}_K B_2$ for $K=\R$ or $\F_2$, where $B_2$ is the link--face incidence matrix. Annual figures use exact elimination over $\F_2$. The 2022 dimension is also 76 over the reals, by the rank of the check matrix.

\textbf{Tax treaties.} The OECD file reports withholding rates by source and counterpart economy for dividends, interest and royalties; we use 2026. For each income type we form the routing gain for all ordered triples of distinct economies with all three rates, and count a triple as inconsistent when the gain exceeds $10^{-9}$. MLI parties come from the OECD list of signatories and parties, matched to the rate file by name (91 of 194 economies).

\textbf{Repair design.} The 2022 complex has 1,771 links, 17,651 triples covered by a common treaty and 23,891 triples whose three links exist, so 6,240 candidates. The rank of the check matrix is 1,616 against a cycle space of dimension 1,692. Because $d_1d_0=0$, the gap equals the $n$th smallest eigenvalue of $d_1^*d_1$ on all links, with $n=80$. Stage one adds the candidate with the largest component outside the row space of the current check matrix. Stage two adds the candidate maximizing the squared inner product of its boundary vector with the current minimum-eigenvalue eigenvector. The random benchmark adds uniformly drawn candidates to the same first-stage design (eight draws). The local diagnostic builds, for each country, the graph of partners joined when all three share a treaty, and requires it to be connected with no isolated partner.

\textbf{Private and social value} (Section~\ref{sec:incentives}). The starting architecture is the 2022 complex plus the 76 first-stage additions, for which $\Gamma=0.0035$. We compute $H^{-1}$ on $C$ from the eigendecomposition of $d_1^*d_1$, dropping its $n-1=79$ zero eigenvalues, which correspond to $\operatorname{im}d_0$. For each of the 6,164 remaining candidates, $u=H^{-1}b$, and $S(\tau)$ and $s(\tau)$ follow from Proposition~\ref{prop:wedge}. The first-order gain in $\Gamma$ is $(\varphi_1^\top b)^2$. In the formation exercise, $H^{-1}$ is updated after each addition by the Sherman--Morrison formula, and $\Gamma$ is recomputed at the end. Clustering is three times the number of triangles in the bilateral graph divided by the number of connected triples.

\textbf{Reproducibility.} The replication package contains a script for each table, the processed membership panel and the hand-built inputs (country list, regional mapping, UK withdrawal dates, MLI party list). The raw DESTA and OECD files come from their providers and are downloaded as described in the README.

\textbf{Accessions} (Section~\ref{sec:accession}). An economy accedes to a base treaty in the first year in which a dyad linking it to the treaty enters into force, if that year is later than the treaty's first year in force. For each country-agreement-year in the risk set, the block of added checks consists of the triples $\{x,i,j\}$, with $x$ the newcomer and $i,j$ incumbents, whose three links exist at $t$ and that no treaty checks at $t$. With $M$ the pseudo-inverse of $H$ on its visible directions, $B$ the matrix of boundary vectors of the block, $U=MB$ and $G=B^\top MB$, the fall in the variance on link $e$ is $[U(I+G)^{-1}U^\top]_{ee}$, which reduces to the Sherman--Morrison formula for a single check. The fall in exposure of country $i$ sums this over its links. An accession makes an invisible inconsistency visible if $B$ has a nonzero component in the null space of $H$ on $C$; the dimension of that null space equals $\dim H^1$ computed over $\F_2$ in every year. The triple-level panel described at the end of this appendix uses the same construction with one check at a time.

\textbf{Accession robustness} (Section~\ref{sec:accession}). All estimates use column (1) of Table~\ref{tab:accession} with standard errors clustered by newcomer and year. (i) Split by whether the agreement has an EU incumbent at $t$: with EU incumbents, 1.49 (0.42) on the log members' gain and $-0.96$ (0.27) on the log outsiders' gain (100 accessions, 31 decisions); without EU incumbents, a single accession has a nonempty block of checks, so no estimate is possible. (ii) Gains measured at $t-2$ for an accession at $t+1$: 1.05 (0.42) and $-0.73$ (0.31), with 78 accessions (28 decisions) for which the lagged risk-set observation exists. (iii) Placebo: the year-$t$ network used to compute gains is relabeled by a fixed random permutation of the 80 economies (seeds 1 to 20), while the risk set and outcomes are as observed. Across the 20 permutations the $z$-statistic on the log members' gain has mean $-0.21$ and standard deviation 1.23, and that on the log outsiders' gain has mean 0.29 and standard deviation 0.77; no permutation reaches the observed values of 3.67 and $-3.65$. (iv) Link-preserving null: for each year the links are kept as observed and the set of checked triangles is replaced by a uniform random set of the same size drawn from all triangles of the link graph (seeds 1 to 20). Checked triangles are 74 to 82 percent of the graph's triangles, so the null randomizes where the unchecked triangles lie, which is what the gains depend on. Each draw is fitted on the observed risk set and outcomes, restricted to rows with a nonempty block of checks under the draw. The $z$-statistic on the log members' gain has mean $-1.20$, standard deviation 0.83 and maximum 0.29; that on the log outsiders' gain has mean $-2.22$, standard deviation 1.06 and minimum $-4.21$, and 2 of 20 draws are at or below the observed $-3.65$. Replacing each gain by its higher-order component, the observed log gain minus its mean across the 20 draws, gives 1.36 (0.38) and $-0.82$ (0.27); the observed and link-implied log members' gains have correlation 0.32. In the planner test of column (4) of Table~\ref{tab:accession}, the $z$-statistic on the log outsiders' share has mean $-2.15$ and minimum $-3.96$ across draws, against $-4.28$ observed; the higher-order components give 0.50 (0.24) on the log social value and $-0.85$ (0.22) on the log outsiders' share. Scripts \texttt{16\_accession\_panel.py} (with \texttt{PERM\_SEED} and \texttt{TRI\_SEED}) and \texttt{20\_accession\_robustness.py}.

\textbf{Repair design} (Section~\ref{sec:repair}). Adding a check raises the rank of $D$ by at most one, so any repair needs at least $\dim\H^1=76$ additions, and the 6,240 candidates contain 76 that achieve it. The first stage adds, one at a time, the candidate least explained by the checks in place (the largest residual after projection onto the row space of $D$), until no uncovered cycle remains. The second stage adds the candidate with the largest squared component along the weakest direction, which to first order raises $\Gamma$ the most. This is a greedy procedure guided by the theory, not an optimum. The random benchmark is the mean of eight draws added to the same first-stage design. After 86 additions the targeted gap is 0.042 against 0.013 for random additions. Of the 126 targeted arrangements, 71 percent involve no EU member; the most frequent participants are Switzerland, Canada, the United States, Costa Rica, Singapore, Jordan and Japan. The local test of Proposition~\ref{prop:local} agrees: only 44 percent of the 80 countries are $\tfrac12$-sound in 2022, and between 40 and 48 percent in every year sampled since 2000; those that pass are mostly EU members, and the rest hold 91 percent of the slots in the targeted additions. Applying the local rule directly takes 2,295 additions, after which 73 countries pass and $\Gamma=0.53$. The expander of Theorem~\ref{thm:3} has about 146 partners per participant and is a benchmark, not a template.

\textbf{Annual series} (Section~\ref{sec:repair}).

\begin{table}[h]
\centering\small
\caption{International treaty membership network, annual series 2016--2022}\label{S-tab:memfull}
\setlength{\tabcolsep}{5pt}\begin{tabular}{lrrrrrr}
\toprule
& Active & Bilateral & Treaty & & Uncovered & Graph\\
Year & treaties & links & triangles & Clustering & cycles ($\F_2$) & gap\\
\midrule
2016 & 370 & 1,575 & 15,504 & 0.817 & 73 & 0.331\\
2017 & 377 & 1,631 & 16,260 & 0.819 & 75 & 0.339\\
2018 & 379 & 1,646 & 16,380 & 0.814 & 67 & 0.356\\
2019 & 386 & 1,718 & 17,220 & 0.816 & 70 & 0.392\\
2020 & 395 & 1,763 & 17,557 & 0.809 & 74 & 0.411\\
2021 & 404 & 1,767 & 17,557 & 0.810 & 78 & 0.411\\
2022 & 410 & 1,771 & 17,651 & 0.811 & 76 & 0.411\\
\bottomrule
\end{tabular}

\smallskip
\parbox{0.92\linewidth}{\footnotesize\emph{Notes:} 80 matched economies. Clustering is the global clustering coefficient of the bilateral graph (three times the number of triangles divided by the number of connected triples). Graph gaps use the unweighted normalized bilateral graph Laplacian. Triangles are filled only with common individual-treaty membership. All annual first-cohomology dimensions are over $\F_2$; the 2022 dimension is also 76 over the reals. Active treaties are base treaties with at least two of the 80 economies as members.}
\end{table}

\textbf{Accession design details} (Section~\ref{sec:accession}). Agreements enter the risk set if they are in force with at least two incumbents among the 80 economies and admit at least one new member in the data. The median newcomer, incumbent and outsider gains are 0.14, 0.31 and 0.09 exposure units, and members' and outsiders' gains have correlation 0.84 in logs. The same test across single candidate triples is much weaker, because private and social values are almost collinear across triples (correlation 0.993 in logs): for 52,348 candidate-years from 1990 to 2021, formation within three years, with year and member-country fixed effects, has coefficient 1.35 (standard error 0.46) on the log private value and $-1.42$ (0.47) on the log social value, with standard errors robust to correlation between triples that share a country; within one year the signs are the same but the estimates are not significant.

\textbf{Members' shares by decile} (Section~\ref{sec:incentives}).

\begin{table}[h]
\centering\small
\caption{Members' share of the value of a candidate check, 2022}\label{S-tab:shares}
\begin{tabular}{ccrc}
\toprule
Decile of & Mean first-order & Median social & Median members'\\
gain in $\Gamma$ & gain in $\Gamma$ & value $S$ & share $s$ (IQR)\\
\midrule
1 (lowest) & $<0.0001$ & 0.12 & 0.58 (0.47--0.62)\\
2 & $<0.0001$ & 0.24 & 0.51 (0.49--0.53)\\
3 & $<0.0001$ & 0.12 & 0.41 (0.40--0.51)\\
4 & $<0.0001$ & 0.16 & 0.48 (0.42--0.52)\\
5 & $<0.0001$ & 0.16 & 0.49 (0.40--0.51)\\
6 & $<0.0001$ & 0.33 & 0.33 (0.25--0.40)\\
7 & 0.001 & 69 & 0.21 (0.17--0.24)\\
8 & 0.010 & 264 & 0.16 (0.13--0.18)\\
9 & 0.016 & 326 & 0.15 (0.13--0.16)\\
10 (highest) & 0.027 & 375 & 0.16 (0.15--0.17)\\
\bottomrule
\end{tabular}

\smallskip
\parbox{0.92\linewidth}{\footnotesize\emph{Notes:} 6,164 candidate checks on the 2022 network after the 76 first-stage additions of Table~\ref{tab:repair}. The first-order gain in $\Gamma$ is the squared component of the candidate's boundary vector along the weakest direction. Social value $S$ is the reduction in total exposure (Proposition~\ref{prop:wedge}), in units of $\sigma^2/(2\kappa)$; total exposure at the starting point is 1,571, or 19.6 per country. The members' share $s$ is defined in~\eqref{eq:share}. IQR: interquartile range.}
\end{table}

\section{Tax treaties: bilateral rates and the missing third check}\label{S-app:tax}

We use the OECD database of treaty withholding-tax rates and the OECD list of parties to the Multilateral Instrument \citep{oecd2026a,oecd2026b}.

A tax treaty lowers the withholding tax on dividends, interest and royalties paid to residents of the partner. Treaties $ij$ and $jk$ raise a question neither answers: can a payment from $i$ to $k$ be made cheaper by passing through $j$? Let $t_{ij}$ be the rate on payments from $i$ to residents of $j$, and $a_{ij}=-\log(1-t_{ij})$. The three treaties are consistent for the triple when
\begin{equation}\label{S-eq:tax}
a_{ik}\le a_{ij}+a_{jk}.
\end{equation}
The \emph{routing gain} $(1-t_{ij})(1-t_{jk})-(1-t_{ik})$ is positive exactly when \eqref{S-eq:tax} fails. This is treaty shopping. It is a three-party check in the sense of Section~\ref{sec:framework}: the discrepancy can be read only from the three rates together. Condition \eqref{S-eq:tax} is a one-sided version of the additive restriction.

\begin{table}[h]
\centering\small
\caption{Routing inconsistency in bilateral withholding-tax treaties, 2026}\label{S-tab:tax}
\resizebox{\linewidth}{!}{\begin{tabular}{lrrr}
\toprule
 & Dividends & Interest & Royalties\\
\midrule
Triples with all three bilateral treaties & 189,694 & 190,436 & 189,536\\
Share where the detour is cheaper & 14.5\% & 8.7\% & 9.3\%\\
Mean gain when positive (percentage points) & 4.4 & 5.5 & 4.4\\
Share through the ten most frequent intermediaries & 30\% & 39\% & 44\%\\
All three countries MLI parties: share of these triples & 72\% & 73\% & 75\%\\
All three countries MLI parties: share of all triples with three treaties & 71\% & 71\% & 72\%\\
\bottomrule
\end{tabular}}

\smallskip
\parbox{0.92\linewidth}{\footnotesize\emph{Notes:} OECD treaty withholding-tax file (194 economies; the MLI party list covers 91 of them). A triple counts when the three rates exist and the economies are distinct; the detour is cheaper when $(1-t_{ij})(1-t_{jk})>1-t_{ik}$. Rates are as reported, one per treaty and income type, without conditioning on ownership thresholds. Triples without a direct treaty are excluded.}
\end{table}

Among triples whose three bilateral links all exist, a detour is cheaper in 9 to 15 percent of cases, by 4 to 5.5 percentage points of the payment on average. It is not confined to a few havens: 100 to 107 economies act as the cheaper intermediary in at least one triple, though the ten most frequent account for 30 to 44 percent. A bilateral link in every pair does not make the system consistent, because consistency is a property of triples. The Multilateral Instrument overlays checks on many bilateral treaties at once. Positive gains are about as common among its parties as elsewhere, because the statutory rates in Table~\ref{S-tab:tax} do not change.

\section{Tax treaties under the Multilateral Instrument}\label{S-app:mli}

The theory predicts that checks are missing where weaknesses are global, not where they are local. Bilateral tax treaties provide the contrast. Under the OECD's Multilateral Instrument (MLI), each jurisdiction lists which of its bilateral tax treaties should be modified by common anti-abuse rules, and a treaty is covered when both partners list it \citep{oecdmli2026}. Treat a triangle of treaties as checked when all three are covered. Among the 80 economies, the 1,679 bilateral tax treaties then form a complex whose covered part is strong, with $\Gamma=1.99$ on the directions the checks see, but which has 420 invisible directions, roughly one for each of the 432 uncovered treaties. These directions are local: a median 95 percent of the mass of the direction that covering a treaty would make visible sits on that treaty itself. By Remark~\ref{rem:invisible}, the two partners then bear about 98 percent of the exposure growing along it, and everyone else about 2 percent. The model predicts that coverage is privately attractive, and it is: 89 percent of 3,207 listing decisions list the treaty. What separates listed from unlisted treaties is reciprocity (96 percent are listed when the partner listed them, 44 percent otherwise), not spillovers to outsiders, which are too small to matter. Local gaps are repaired privately; the global gaps in trade agreements are not.

\textbf{Data.} The network of bilateral tax treaties among the 80 economies is taken from the OECD treaty withholding-tax data used in Appendix~\ref{S-app:tax}: two economies are linked if either reports a treaty rate for the other (1,679 treaties). Listing decisions and coverage come from the Annex of the Excel version of the OECD BEPS MLI Matching Database, status as of 12 November 2026 \citep{oecdmli2026}: a jurisdiction \emph{lists} a treaty if it appears among that jurisdiction's notified agreements, and a treaty is \emph{covered} if it appears among the matched agreements. Jurisdiction names are mapped to ISO codes through the names in the withholding-tax data; regional agreements and jurisdictions outside the 80 are dropped. Sixty-nine of the 80 economies are MLI jurisdictions, giving 3,207 listing decisions (a jurisdiction and one of its treaties with another of the 80); 1,247 treaties are covered.

\textbf{Construction.} A triangle is checked when its three treaties are covered (12,815 triangles). For a treaty $e=ik$, its block is the set of triangles $\{i,k,x\}$ whose other two treaties are covered. For an uncovered treaty, the block's projection onto $\ker H$ spans the invisible directions that covering $e$ would make visible; for a covered treaty, the directions that would become invisible without the block are those on which the block's Gram matrix $B^\top H^+B$ has eigenvalue one. Let $W$ be an orthonormal basis of these directions. Following Remark~\ref{rem:invisible}, exposure along them grows at a rate proportional to their mass $\sum_{e'\ni j}\sum_w w_{e'}^2$ on each country $j$'s links, and we report each country's share of the total. In 2,972 of the 3,207 decisions the block is nonempty and exactly one direction is involved in almost all of them.

\textbf{Results.} The median listing country's share is 0.49, the partner's is 0.49 and everyone else's is 0.02. The direction is concentrated on the treaty itself (median mass 0.95). In a logit of listing with listing-country fixed effects, among the 60 jurisdictions that list some but not all of their treaties with a nonempty block (2,627 decisions, 331 not listed), the log share of third parties and outsiders has coefficient $-1.38$ (standard error 0.30, clustered by jurisdiction) on its own, $-0.43$ (0.49) after controlling for the log number of checks in the block, the age of the treaty and the partner's degree, and turns positive once the partner's own listing is added; the partner's listing has coefficient 5.70 (0.58). Script \texttt{19\_mli\_panel.py} produces these estimates and the figures above.

\bibliographystyle{plainnat}
\bibliography{references}

\end{document}